\documentclass[twocolumn,showpacs,prc,amsmath,amssymb,superscriptaddress]{revtex4}

\usepackage{graphicx} 
\usepackage{dcolumn}  
\usepackage{bm}       
\usepackage{color}
\usepackage{epsfig}
\usepackage{fancyhdr}
\usepackage{amsmath}
\usepackage{amsfonts}
  \definecolor{snow}{rgb}{1.000000,0.980392,0.980392}
  \definecolor{ghost white}{rgb}{0.972549,0.972549,1.000000}
  \definecolor{GhostWhite}{rgb}{0.972549,0.972549,1.000000}
  \definecolor{white smoke}{rgb}{0.960784,0.960784,0.960784}
  \definecolor{WhiteSmoke}{rgb}{0.960784,0.960784,0.960784}
  \definecolor{gainsboro}{rgb}{0.862745,0.862745,0.862745}
  \definecolor{floral white}{rgb}{1.000000,0.980392,0.941176}
  \definecolor{FloralWhite}{rgb}{1.000000,0.980392,0.941176}
  \definecolor{old lace}{rgb}{0.992157,0.960784,0.901961}
  \definecolor{OldLace}{rgb}{0.992157,0.960784,0.901961}
  \definecolor{linen}{rgb}{0.980392,0.941176,0.901961}
  \definecolor{antique white}{rgb}{0.980392,0.921569,0.843137}
  \definecolor{AntiqueWhite}{rgb}{0.980392,0.921569,0.843137}
  \definecolor{papaya whip}{rgb}{1.000000,0.937255,0.835294}
  \definecolor{PapayaWhip}{rgb}{1.000000,0.937255,0.835294}
  \definecolor{blanched almond}{rgb}{1.000000,0.921569,0.803922}
  \definecolor{BlanchedAlmond}{rgb}{1.000000,0.921569,0.803922}
  \definecolor{bisque}{rgb}{1.000000,0.894118,0.768627}
  \definecolor{peach puff}{rgb}{1.000000,0.854902,0.725490}
  \definecolor{PeachPuff}{rgb}{1.000000,0.854902,0.725490}
  \definecolor{navajo white}{rgb}{1.000000,0.870588,0.678431}
  \definecolor{NavajoWhite}{rgb}{1.000000,0.870588,0.678431}
  \definecolor{moccasin}{rgb}{1.000000,0.894118,0.709804}
  \definecolor{cornsilk}{rgb}{1.000000,0.972549,0.862745}
  \definecolor{ivory}{rgb}{1.000000,1.000000,0.941176}
  \definecolor{lemon chiffon}{rgb}{1.000000,0.980392,0.803922}
  \definecolor{LemonChiffon}{rgb}{1.000000,0.980392,0.803922}
  \definecolor{seashell}{rgb}{1.000000,0.960784,0.933333}
  \definecolor{honeydew}{rgb}{0.941176,1.000000,0.941176}
  \definecolor{mint cream}{rgb}{0.960784,1.000000,0.980392}
  \definecolor{MintCream}{rgb}{0.960784,1.000000,0.980392}
  \definecolor{azure}{rgb}{0.941176,1.000000,1.000000}
  \definecolor{alice blue}{rgb}{0.941176,0.972549,1.000000}
  \definecolor{AliceBlue}{rgb}{0.941176,0.972549,1.000000}
  \definecolor{lavender}{rgb}{0.901961,0.901961,0.980392}
  \definecolor{lavender blush}{rgb}{1.000000,0.941176,0.960784}
  \definecolor{LavenderBlush}{rgb}{1.000000,0.941176,0.960784}
  \definecolor{misty rose}{rgb}{1.000000,0.894118,0.882353}
  \definecolor{MistyRose}{rgb}{1.000000,0.894118,0.882353}
  \definecolor{white}{rgb}{1.000000,1.000000,1.000000}
  \definecolor{black}{rgb}{0.000000,0.000000,0.000000}
  \definecolor{dark slate gray}{rgb}{0.184314,0.309804,0.309804}
  \definecolor{DarkSlateGray}{rgb}{0.184314,0.309804,0.309804}
  \definecolor{dark slate grey}{rgb}{0.184314,0.309804,0.309804}
  \definecolor{DarkSlateGrey}{rgb}{0.184314,0.309804,0.309804}
  \definecolor{dim gray}{rgb}{0.411765,0.411765,0.411765}
  \definecolor{DimGray}{rgb}{0.411765,0.411765,0.411765}
  \definecolor{dim grey}{rgb}{0.411765,0.411765,0.411765}
  \definecolor{DimGrey}{rgb}{0.411765,0.411765,0.411765}
  \definecolor{slate gray}{rgb}{0.439216,0.501961,0.564706}
  \definecolor{SlateGray}{rgb}{0.439216,0.501961,0.564706}
  \definecolor{slate grey}{rgb}{0.439216,0.501961,0.564706}
  \definecolor{SlateGrey}{rgb}{0.439216,0.501961,0.564706}
  \definecolor{light slate gray}{rgb}{0.466667,0.533333,0.600000}
  \definecolor{LightSlateGray}{rgb}{0.466667,0.533333,0.600000}
  \definecolor{light slate grey}{rgb}{0.466667,0.533333,0.600000}
  \definecolor{LightSlateGrey}{rgb}{0.466667,0.533333,0.600000}
  \definecolor{gray}{rgb}{0.745098,0.745098,0.745098}
  \definecolor{grey}{rgb}{0.745098,0.745098,0.745098}
  \definecolor{light grey}{rgb}{0.827451,0.827451,0.827451}
  \definecolor{LightGrey}{rgb}{0.827451,0.827451,0.827451}
  \definecolor{light gray}{rgb}{0.827451,0.827451,0.827451}
  \definecolor{LightGray}{rgb}{0.827451,0.827451,0.827451}
  \definecolor{midnight blue}{rgb}{0.098039,0.098039,0.439216}
  \definecolor{MidnightBlue}{rgb}{0.098039,0.098039,0.439216}
  \definecolor{navy}{rgb}{0.000000,0.000000,0.501961}
  \definecolor{navy blue}{rgb}{0.000000,0.000000,0.501961}
  \definecolor{NavyBlue}{rgb}{0.000000,0.000000,0.501961}
  \definecolor{cornflower blue}{rgb}{0.392157,0.584314,0.929412}
  \definecolor{CornflowerBlue}{rgb}{0.392157,0.584314,0.929412}
  \definecolor{dark slate blue}{rgb}{0.282353,0.239216,0.545098}
  \definecolor{DarkSlateBlue}{rgb}{0.282353,0.239216,0.545098}
  \definecolor{slate blue}{rgb}{0.415686,0.352941,0.803922}
  \definecolor{SlateBlue}{rgb}{0.415686,0.352941,0.803922}
  \definecolor{medium slate blue}{rgb}{0.482353,0.407843,0.933333}
  \definecolor{MediumSlateBlue}{rgb}{0.482353,0.407843,0.933333}
  \definecolor{light slate blue}{rgb}{0.517647,0.439216,1.000000}
  \definecolor{LightSlateBlue}{rgb}{0.517647,0.439216,1.000000}
  \definecolor{medium blue}{rgb}{0.000000,0.000000,0.803922}
  \definecolor{MediumBlue}{rgb}{0.000000,0.000000,0.803922}
  \definecolor{royal blue}{rgb}{0.254902,0.411765,0.882353}
  \definecolor{RoyalBlue}{rgb}{0.254902,0.411765,0.882353}
  \definecolor{blue}{rgb}{0.000000,0.000000,1.000000}
  \definecolor{dodger blue}{rgb}{0.117647,0.564706,1.000000}
  \definecolor{DodgerBlue}{rgb}{0.117647,0.564706,1.000000}
  \definecolor{deep sky blue}{rgb}{0.000000,0.749020,1.000000}
  \definecolor{DeepSkyBlue}{rgb}{0.000000,0.749020,1.000000}
  \definecolor{sky blue}{rgb}{0.529412,0.807843,0.921569}
  \definecolor{SkyBlue}{rgb}{0.529412,0.807843,0.921569}
  \definecolor{light sky blue}{rgb}{0.529412,0.807843,0.980392}
  \definecolor{LightSkyBlue}{rgb}{0.529412,0.807843,0.980392}
  \definecolor{steel blue}{rgb}{0.274510,0.509804,0.705882}
  \definecolor{SteelBlue}{rgb}{0.274510,0.509804,0.705882}
  \definecolor{light steel blue}{rgb}{0.690196,0.768627,0.870588}
  \definecolor{LightSteelBlue}{rgb}{0.690196,0.768627,0.870588}
  \definecolor{light blue}{rgb}{0.678431,0.847059,0.901961}
  \definecolor{LightBlue}{rgb}{0.678431,0.847059,0.901961}
  \definecolor{powder blue}{rgb}{0.690196,0.878431,0.901961}
  \definecolor{PowderBlue}{rgb}{0.690196,0.878431,0.901961}
  \definecolor{pale turquoise}{rgb}{0.686275,0.933333,0.933333}
  \definecolor{PaleTurquoise}{rgb}{0.686275,0.933333,0.933333}
  \definecolor{dark turquoise}{rgb}{0.000000,0.807843,0.819608}
  \definecolor{DarkTurquoise}{rgb}{0.000000,0.807843,0.819608}
  \definecolor{medium turquoise}{rgb}{0.282353,0.819608,0.800000}
  \definecolor{MediumTurquoise}{rgb}{0.282353,0.819608,0.800000}
  \definecolor{turquoise}{rgb}{0.250980,0.878431,0.815686}
  \definecolor{cyan}{rgb}{0.000000,1.000000,1.000000}
  \definecolor{light cyan}{rgb}{0.878431,1.000000,1.000000}
  \definecolor{LightCyan}{rgb}{0.878431,1.000000,1.000000}
  \definecolor{cadet blue}{rgb}{0.372549,0.619608,0.627451}
  \definecolor{CadetBlue}{rgb}{0.372549,0.619608,0.627451}
  \definecolor{medium aquamarine}{rgb}{0.400000,0.803922,0.666667}
  \definecolor{MediumAquamarine}{rgb}{0.400000,0.803922,0.666667}
  \definecolor{aquamarine}{rgb}{0.498039,1.000000,0.831373}
  \definecolor{dark green}{rgb}{0.000000,0.392157,0.000000}
  \definecolor{DarkGreen}{rgb}{0.000000,0.392157,0.000000}
  \definecolor{dark olive green}{rgb}{0.333333,0.419608,0.184314}
  \definecolor{DarkOliveGreen}{rgb}{0.333333,0.419608,0.184314}
  \definecolor{dark sea green}{rgb}{0.560784,0.737255,0.560784}
  \definecolor{DarkSeaGreen}{rgb}{0.560784,0.737255,0.560784}
  \definecolor{sea green}{rgb}{0.180392,0.545098,0.341176}
  \definecolor{SeaGreen}{rgb}{0.180392,0.545098,0.341176}
  \definecolor{medium sea green}{rgb}{0.235294,0.701961,0.443137}
  \definecolor{MediumSeaGreen}{rgb}{0.235294,0.701961,0.443137}
  \definecolor{light sea green}{rgb}{0.125490,0.698039,0.666667}
  \definecolor{LightSeaGreen}{rgb}{0.125490,0.698039,0.666667}
  \definecolor{pale green}{rgb}{0.596078,0.984314,0.596078}
  \definecolor{PaleGreen}{rgb}{0.596078,0.984314,0.596078}
  \definecolor{spring green}{rgb}{0.000000,1.000000,0.498039}
  \definecolor{SpringGreen}{rgb}{0.000000,1.000000,0.498039}
  \definecolor{lawn green}{rgb}{0.486275,0.988235,0.000000}
  \definecolor{LawnGreen}{rgb}{0.486275,0.988235,0.000000}
  \definecolor{green}{rgb}{0.000000,1.000000,0.000000}
  \definecolor{chartreuse}{rgb}{0.498039,1.000000,0.000000}
  \definecolor{medium spring green}{rgb}{0.000000,0.980392,0.603922}
  \definecolor{MediumSpringGreen}{rgb}{0.000000,0.980392,0.603922}
  \definecolor{green yellow}{rgb}{0.678431,1.000000,0.184314}
  \definecolor{GreenYellow}{rgb}{0.678431,1.000000,0.184314}
  \definecolor{lime green}{rgb}{0.196078,0.803922,0.196078}
  \definecolor{LimeGreen}{rgb}{0.196078,0.803922,0.196078}
  \definecolor{yellow green}{rgb}{0.603922,0.803922,0.196078}
  \definecolor{YellowGreen}{rgb}{0.603922,0.803922,0.196078}
  \definecolor{forest green}{rgb}{0.133333,0.545098,0.133333}
  \definecolor{ForestGreen}{rgb}{0.133333,0.545098,0.133333}
  \definecolor{olive drab}{rgb}{0.419608,0.556863,0.137255}
  \definecolor{OliveDrab}{rgb}{0.419608,0.556863,0.137255}
  \definecolor{dark khaki}{rgb}{0.741176,0.717647,0.419608}
  \definecolor{DarkKhaki}{rgb}{0.741176,0.717647,0.419608}
  \definecolor{khaki}{rgb}{0.941176,0.901961,0.549020}
  \definecolor{pale goldenrod}{rgb}{0.933333,0.909804,0.666667}
  \definecolor{PaleGoldenrod}{rgb}{0.933333,0.909804,0.666667}
  \definecolor{light goldenrod yellow}{rgb}{0.980392,0.980392,0.823529}
  \definecolor{LightGoldenrodYellow}{rgb}{0.980392,0.980392,0.823529}
  \definecolor{light yellow}{rgb}{1.000000,1.000000,0.878431}
  \definecolor{LightYellow}{rgb}{1.000000,1.000000,0.878431}
  \definecolor{yellow}{rgb}{1.000000,1.000000,0.000000}
  \definecolor{gold}{rgb}{1.000000,0.843137,0.000000}
  \definecolor{light goldenrod}{rgb}{0.933333,0.866667,0.509804}
  \definecolor{LightGoldenrod}{rgb}{0.933333,0.866667,0.509804}
  \definecolor{goldenrod}{rgb}{0.854902,0.647059,0.125490}
  \definecolor{dark goldenrod}{rgb}{0.721569,0.525490,0.043137}
  \definecolor{DarkGoldenrod}{rgb}{0.721569,0.525490,0.043137}
  \definecolor{rosy brown}{rgb}{0.737255,0.560784,0.560784}
  \definecolor{RosyBrown}{rgb}{0.737255,0.560784,0.560784}
  \definecolor{indian red}{rgb}{0.803922,0.360784,0.360784}
  \definecolor{IndianRed}{rgb}{0.803922,0.360784,0.360784}
  \definecolor{saddle brown}{rgb}{0.545098,0.270588,0.074510}
  \definecolor{SaddleBrown}{rgb}{0.545098,0.270588,0.074510}
  \definecolor{sienna}{rgb}{0.627451,0.321569,0.176471}
  \definecolor{peru}{rgb}{0.803922,0.521569,0.247059}
  \definecolor{burlywood}{rgb}{0.870588,0.721569,0.529412}
  \definecolor{beige}{rgb}{0.960784,0.960784,0.862745}
  \definecolor{wheat}{rgb}{0.960784,0.870588,0.701961}
  \definecolor{sandy brown}{rgb}{0.956863,0.643137,0.376471}
  \definecolor{SandyBrown}{rgb}{0.956863,0.643137,0.376471}
  \definecolor{tan}{rgb}{0.823529,0.705882,0.549020}
  \definecolor{chocolate}{rgb}{0.823529,0.411765,0.117647}
  \definecolor{firebrick}{rgb}{0.698039,0.133333,0.133333}
  \definecolor{brown}{rgb}{0.647059,0.164706,0.164706}
  \definecolor{dark salmon}{rgb}{0.913725,0.588235,0.478431}
  \definecolor{DarkSalmon}{rgb}{0.913725,0.588235,0.478431}
  \definecolor{salmon}{rgb}{0.980392,0.501961,0.447059}
  \definecolor{light salmon}{rgb}{1.000000,0.627451,0.478431}
  \definecolor{LightSalmon}{rgb}{1.000000,0.627451,0.478431}
  \definecolor{orange}{rgb}{1.000000,0.647059,0.000000}
  \definecolor{dark orange}{rgb}{1.000000,0.549020,0.000000}
  \definecolor{DarkOrange}{rgb}{1.000000,0.549020,0.000000}
  \definecolor{coral}{rgb}{1.000000,0.498039,0.313726}
  \definecolor{light coral}{rgb}{0.941176,0.501961,0.501961}
  \definecolor{LightCoral}{rgb}{0.941176,0.501961,0.501961}
  \definecolor{tomato}{rgb}{1.000000,0.388235,0.278431}
  \definecolor{orange red}{rgb}{1.000000,0.270588,0.000000}
  \definecolor{OrangeRed}{rgb}{1.000000,0.270588,0.000000}
  \definecolor{red}{rgb}{1.000000,0.000000,0.000000}
  \definecolor{hot pink}{rgb}{1.000000,0.411765,0.705882}
  \definecolor{HotPink}{rgb}{1.000000,0.411765,0.705882}
  \definecolor{deep pink}{rgb}{1.000000,0.078431,0.576471}
  \definecolor{DeepPink}{rgb}{1.000000,0.078431,0.576471}
  \definecolor{pink}{rgb}{1.000000,0.752941,0.796078}
  \definecolor{light pink}{rgb}{1.000000,0.713726,0.756863}
  \definecolor{LightPink}{rgb}{1.000000,0.713726,0.756863}
  \definecolor{pale violet red}{rgb}{0.858824,0.439216,0.576471}
  \definecolor{PaleVioletRed}{rgb}{0.858824,0.439216,0.576471}
  \definecolor{maroon}{rgb}{0.690196,0.188235,0.376471}
  \definecolor{medium violet red}{rgb}{0.780392,0.082353,0.521569}
  \definecolor{MediumVioletRed}{rgb}{0.780392,0.082353,0.521569}
  \definecolor{violet red}{rgb}{0.815686,0.125490,0.564706}
  \definecolor{VioletRed}{rgb}{0.815686,0.125490,0.564706}
  \definecolor{magenta}{rgb}{1.000000,0.000000,1.000000}
  \definecolor{violet}{rgb}{0.933333,0.509804,0.933333}
  \definecolor{plum}{rgb}{0.866667,0.627451,0.866667}
  \definecolor{orchid}{rgb}{0.854902,0.439216,0.839216}
  \definecolor{medium orchid}{rgb}{0.729412,0.333333,0.827451}
  \definecolor{MediumOrchid}{rgb}{0.729412,0.333333,0.827451}
  \definecolor{dark orchid}{rgb}{0.600000,0.196078,0.800000}
  \definecolor{DarkOrchid}{rgb}{0.600000,0.196078,0.800000}
  \definecolor{dark violet}{rgb}{0.580392,0.000000,0.827451}
  \definecolor{DarkViolet}{rgb}{0.580392,0.000000,0.827451}
  \definecolor{blue violet}{rgb}{0.541176,0.168627,0.886275}
  \definecolor{BlueViolet}{rgb}{0.541176,0.168627,0.886275}
  \definecolor{purple}{rgb}{0.627451,0.125490,0.941176}
  \definecolor{medium purple}{rgb}{0.576471,0.439216,0.858824}
  \definecolor{MediumPurple}{rgb}{0.576471,0.439216,0.858824}
  \definecolor{thistle}{rgb}{0.847059,0.749020,0.847059}
  \definecolor{snow1}{rgb}{1.000000,0.980392,0.980392}
  \definecolor{snow2}{rgb}{0.933333,0.913725,0.913725}
  \definecolor{snow3}{rgb}{0.803922,0.788235,0.788235}
  \definecolor{snow4}{rgb}{0.545098,0.537255,0.537255}
  \definecolor{seashell1}{rgb}{1.000000,0.960784,0.933333}
  \definecolor{seashell2}{rgb}{0.933333,0.898039,0.870588}
  \definecolor{seashell3}{rgb}{0.803922,0.772549,0.749020}
  \definecolor{seashell4}{rgb}{0.545098,0.525490,0.509804}
  \definecolor{AntiqueWhite1}{rgb}{1.000000,0.937255,0.858824}
  \definecolor{AntiqueWhite2}{rgb}{0.933333,0.874510,0.800000}
  \definecolor{AntiqueWhite3}{rgb}{0.803922,0.752941,0.690196}
  \definecolor{AntiqueWhite4}{rgb}{0.545098,0.513726,0.470588}
  \definecolor{bisque1}{rgb}{1.000000,0.894118,0.768627}
  \definecolor{bisque2}{rgb}{0.933333,0.835294,0.717647}
  \definecolor{bisque3}{rgb}{0.803922,0.717647,0.619608}
  \definecolor{bisque4}{rgb}{0.545098,0.490196,0.419608}
  \definecolor{PeachPuff1}{rgb}{1.000000,0.854902,0.725490}
  \definecolor{PeachPuff2}{rgb}{0.933333,0.796078,0.678431}
  \definecolor{PeachPuff3}{rgb}{0.803922,0.686275,0.584314}
  \definecolor{PeachPuff4}{rgb}{0.545098,0.466667,0.396078}
  \definecolor{NavajoWhite1}{rgb}{1.000000,0.870588,0.678431}
  \definecolor{NavajoWhite2}{rgb}{0.933333,0.811765,0.631373}
  \definecolor{NavajoWhite3}{rgb}{0.803922,0.701961,0.545098}
  \definecolor{NavajoWhite4}{rgb}{0.545098,0.474510,0.368627}
  \definecolor{LemonChiffon1}{rgb}{1.000000,0.980392,0.803922}
  \definecolor{LemonChiffon2}{rgb}{0.933333,0.913725,0.749020}
  \definecolor{LemonChiffon3}{rgb}{0.803922,0.788235,0.647059}
  \definecolor{LemonChiffon4}{rgb}{0.545098,0.537255,0.439216}
  \definecolor{cornsilk1}{rgb}{1.000000,0.972549,0.862745}
  \definecolor{cornsilk2}{rgb}{0.933333,0.909804,0.803922}
  \definecolor{cornsilk3}{rgb}{0.803922,0.784314,0.694118}
  \definecolor{cornsilk4}{rgb}{0.545098,0.533333,0.470588}
  \definecolor{ivory1}{rgb}{1.000000,1.000000,0.941176}
  \definecolor{ivory2}{rgb}{0.933333,0.933333,0.878431}
  \definecolor{ivory3}{rgb}{0.803922,0.803922,0.756863}
  \definecolor{ivory4}{rgb}{0.545098,0.545098,0.513726}
  \definecolor{honeydew1}{rgb}{0.941176,1.000000,0.941176}
  \definecolor{honeydew2}{rgb}{0.878431,0.933333,0.878431}
  \definecolor{honeydew3}{rgb}{0.756863,0.803922,0.756863}
  \definecolor{honeydew4}{rgb}{0.513726,0.545098,0.513726}
  \definecolor{LavenderBlush1}{rgb}{1.000000,0.941176,0.960784}
  \definecolor{LavenderBlush2}{rgb}{0.933333,0.878431,0.898039}
  \definecolor{LavenderBlush3}{rgb}{0.803922,0.756863,0.772549}
  \definecolor{LavenderBlush4}{rgb}{0.545098,0.513726,0.525490}
  \definecolor{MistyRose1}{rgb}{1.000000,0.894118,0.882353}
  \definecolor{MistyRose2}{rgb}{0.933333,0.835294,0.823529}
  \definecolor{MistyRose3}{rgb}{0.803922,0.717647,0.709804}
  \definecolor{MistyRose4}{rgb}{0.545098,0.490196,0.482353}
  \definecolor{azure1}{rgb}{0.941176,1.000000,1.000000}
  \definecolor{azure2}{rgb}{0.878431,0.933333,0.933333}
  \definecolor{azure3}{rgb}{0.756863,0.803922,0.803922}
  \definecolor{azure4}{rgb}{0.513726,0.545098,0.545098}
  \definecolor{SlateBlue1}{rgb}{0.513726,0.435294,1.000000}
  \definecolor{SlateBlue2}{rgb}{0.478431,0.403922,0.933333}
  \definecolor{SlateBlue3}{rgb}{0.411765,0.349020,0.803922}
  \definecolor{SlateBlue4}{rgb}{0.278431,0.235294,0.545098}
  \definecolor{RoyalBlue1}{rgb}{0.282353,0.462745,1.000000}
  \definecolor{RoyalBlue2}{rgb}{0.262745,0.431373,0.933333}
  \definecolor{RoyalBlue3}{rgb}{0.227451,0.372549,0.803922}
  \definecolor{RoyalBlue4}{rgb}{0.152941,0.250980,0.545098}
  \definecolor{blue1}{rgb}{0.000000,0.000000,1.000000}
  \definecolor{blue2}{rgb}{0.000000,0.000000,0.933333}
  \definecolor{blue3}{rgb}{0.000000,0.000000,0.803922}
  \definecolor{blue4}{rgb}{0.000000,0.000000,0.545098}
  \definecolor{DodgerBlue1}{rgb}{0.117647,0.564706,1.000000}
  \definecolor{DodgerBlue2}{rgb}{0.109804,0.525490,0.933333}
  \definecolor{DodgerBlue3}{rgb}{0.094118,0.454902,0.803922}
  \definecolor{DodgerBlue4}{rgb}{0.062745,0.305882,0.545098}
  \definecolor{SteelBlue1}{rgb}{0.388235,0.721569,1.000000}
  \definecolor{SteelBlue2}{rgb}{0.360784,0.674510,0.933333}
  \definecolor{SteelBlue3}{rgb}{0.309804,0.580392,0.803922}
  \definecolor{SteelBlue4}{rgb}{0.211765,0.392157,0.545098}
  \definecolor{DeepSkyBlue1}{rgb}{0.000000,0.749020,1.000000}
  \definecolor{DeepSkyBlue2}{rgb}{0.000000,0.698039,0.933333}
  \definecolor{DeepSkyBlue3}{rgb}{0.000000,0.603922,0.803922}
  \definecolor{DeepSkyBlue4}{rgb}{0.000000,0.407843,0.545098}
  \definecolor{SkyBlue1}{rgb}{0.529412,0.807843,1.000000}
  \definecolor{SkyBlue2}{rgb}{0.494118,0.752941,0.933333}
  \definecolor{SkyBlue3}{rgb}{0.423529,0.650980,0.803922}
  \definecolor{SkyBlue4}{rgb}{0.290196,0.439216,0.545098}
  \definecolor{LightSkyBlue1}{rgb}{0.690196,0.886275,1.000000}
  \definecolor{LightSkyBlue2}{rgb}{0.643137,0.827451,0.933333}
  \definecolor{LightSkyBlue3}{rgb}{0.552941,0.713726,0.803922}
  \definecolor{LightSkyBlue4}{rgb}{0.376471,0.482353,0.545098}
  \definecolor{SlateGray1}{rgb}{0.776471,0.886275,1.000000}
  \definecolor{SlateGray2}{rgb}{0.725490,0.827451,0.933333}
  \definecolor{SlateGray3}{rgb}{0.623529,0.713726,0.803922}
  \definecolor{SlateGray4}{rgb}{0.423529,0.482353,0.545098}
  \definecolor{LightSteelBlue1}{rgb}{0.792157,0.882353,1.000000}
  \definecolor{LightSteelBlue2}{rgb}{0.737255,0.823529,0.933333}
  \definecolor{LightSteelBlue3}{rgb}{0.635294,0.709804,0.803922}
  \definecolor{LightSteelBlue4}{rgb}{0.431373,0.482353,0.545098}
  \definecolor{LightBlue1}{rgb}{0.749020,0.937255,1.000000}
  \definecolor{LightBlue2}{rgb}{0.698039,0.874510,0.933333}
  \definecolor{LightBlue3}{rgb}{0.603922,0.752941,0.803922}
  \definecolor{LightBlue4}{rgb}{0.407843,0.513726,0.545098}
  \definecolor{LightCyan1}{rgb}{0.878431,1.000000,1.000000}
  \definecolor{LightCyan2}{rgb}{0.819608,0.933333,0.933333}
  \definecolor{LightCyan3}{rgb}{0.705882,0.803922,0.803922}
  \definecolor{LightCyan4}{rgb}{0.478431,0.545098,0.545098}
  \definecolor{PaleTurquoise1}{rgb}{0.733333,1.000000,1.000000}
  \definecolor{PaleTurquoise2}{rgb}{0.682353,0.933333,0.933333}
  \definecolor{PaleTurquoise3}{rgb}{0.588235,0.803922,0.803922}
  \definecolor{PaleTurquoise4}{rgb}{0.400000,0.545098,0.545098}
  \definecolor{CadetBlue1}{rgb}{0.596078,0.960784,1.000000}
  \definecolor{CadetBlue2}{rgb}{0.556863,0.898039,0.933333}
  \definecolor{CadetBlue3}{rgb}{0.478431,0.772549,0.803922}
  \definecolor{CadetBlue4}{rgb}{0.325490,0.525490,0.545098}
  \definecolor{turquoise1}{rgb}{0.000000,0.960784,1.000000}
  \definecolor{turquoise2}{rgb}{0.000000,0.898039,0.933333}
  \definecolor{turquoise3}{rgb}{0.000000,0.772549,0.803922}
  \definecolor{turquoise4}{rgb}{0.000000,0.525490,0.545098}
  \definecolor{cyan1}{rgb}{0.000000,1.000000,1.000000}
  \definecolor{cyan2}{rgb}{0.000000,0.933333,0.933333}
  \definecolor{cyan3}{rgb}{0.000000,0.803922,0.803922}
  \definecolor{cyan4}{rgb}{0.000000,0.545098,0.545098}
  \definecolor{DarkSlateGray1}{rgb}{0.592157,1.000000,1.000000}
  \definecolor{DarkSlateGray2}{rgb}{0.552941,0.933333,0.933333}
  \definecolor{DarkSlateGray3}{rgb}{0.474510,0.803922,0.803922}
  \definecolor{DarkSlateGray4}{rgb}{0.321569,0.545098,0.545098}
  \definecolor{aquamarine1}{rgb}{0.498039,1.000000,0.831373}
  \definecolor{aquamarine2}{rgb}{0.462745,0.933333,0.776471}
  \definecolor{aquamarine3}{rgb}{0.400000,0.803922,0.666667}
  \definecolor{aquamarine4}{rgb}{0.270588,0.545098,0.454902}
  \definecolor{DarkSeaGreen1}{rgb}{0.756863,1.000000,0.756863}
  \definecolor{DarkSeaGreen2}{rgb}{0.705882,0.933333,0.705882}
  \definecolor{DarkSeaGreen3}{rgb}{0.607843,0.803922,0.607843}
  \definecolor{DarkSeaGreen4}{rgb}{0.411765,0.545098,0.411765}
  \definecolor{SeaGreen1}{rgb}{0.329412,1.000000,0.623529}
  \definecolor{SeaGreen2}{rgb}{0.305882,0.933333,0.580392}
  \definecolor{SeaGreen3}{rgb}{0.262745,0.803922,0.501961}
  \definecolor{SeaGreen4}{rgb}{0.180392,0.545098,0.341176}
  \definecolor{PaleGreen1}{rgb}{0.603922,1.000000,0.603922}
  \definecolor{PaleGreen2}{rgb}{0.564706,0.933333,0.564706}
  \definecolor{PaleGreen3}{rgb}{0.486275,0.803922,0.486275}
  \definecolor{PaleGreen4}{rgb}{0.329412,0.545098,0.329412}
  \definecolor{SpringGreen1}{rgb}{0.000000,1.000000,0.498039}
  \definecolor{SpringGreen2}{rgb}{0.000000,0.933333,0.462745}
  \definecolor{SpringGreen3}{rgb}{0.000000,0.803922,0.400000}
  \definecolor{SpringGreen4}{rgb}{0.000000,0.545098,0.270588}
  \definecolor{green1}{rgb}{0.000000,1.000000,0.000000}
  \definecolor{green2}{rgb}{0.000000,0.933333,0.000000}
  \definecolor{green3}{rgb}{0.000000,0.803922,0.000000}
  \definecolor{green4}{rgb}{0.000000,0.545098,0.000000}
  \definecolor{chartreuse1}{rgb}{0.498039,1.000000,0.000000}
  \definecolor{chartreuse2}{rgb}{0.462745,0.933333,0.000000}
  \definecolor{chartreuse3}{rgb}{0.400000,0.803922,0.000000}
  \definecolor{chartreuse4}{rgb}{0.270588,0.545098,0.000000}
  \definecolor{OliveDrab1}{rgb}{0.752941,1.000000,0.243137}
  \definecolor{OliveDrab2}{rgb}{0.701961,0.933333,0.227451}
  \definecolor{OliveDrab3}{rgb}{0.603922,0.803922,0.196078}
  \definecolor{OliveDrab4}{rgb}{0.411765,0.545098,0.133333}
  \definecolor{DarkOliveGreen1}{rgb}{0.792157,1.000000,0.439216}
  \definecolor{DarkOliveGreen2}{rgb}{0.737255,0.933333,0.407843}
  \definecolor{DarkOliveGreen3}{rgb}{0.635294,0.803922,0.352941}
  \definecolor{DarkOliveGreen4}{rgb}{0.431373,0.545098,0.239216}
  \definecolor{khaki1}{rgb}{1.000000,0.964706,0.560784}
  \definecolor{khaki2}{rgb}{0.933333,0.901961,0.521569}
  \definecolor{khaki3}{rgb}{0.803922,0.776471,0.450980}
  \definecolor{khaki4}{rgb}{0.545098,0.525490,0.305882}
  \definecolor{LightGoldenrod1}{rgb}{1.000000,0.925490,0.545098}
  \definecolor{LightGoldenrod2}{rgb}{0.933333,0.862745,0.509804}
  \definecolor{LightGoldenrod3}{rgb}{0.803922,0.745098,0.439216}
  \definecolor{LightGoldenrod4}{rgb}{0.545098,0.505882,0.298039}
  \definecolor{LightYellow1}{rgb}{1.000000,1.000000,0.878431}
  \definecolor{LightYellow2}{rgb}{0.933333,0.933333,0.819608}
  \definecolor{LightYellow3}{rgb}{0.803922,0.803922,0.705882}
  \definecolor{LightYellow4}{rgb}{0.545098,0.545098,0.478431}
  \definecolor{yellow1}{rgb}{1.000000,1.000000,0.000000}
  \definecolor{yellow2}{rgb}{0.933333,0.933333,0.000000}
  \definecolor{yellow3}{rgb}{0.803922,0.803922,0.000000}
  \definecolor{yellow4}{rgb}{0.545098,0.545098,0.000000}
  \definecolor{gold1}{rgb}{1.000000,0.843137,0.000000}
  \definecolor{gold2}{rgb}{0.933333,0.788235,0.000000}
  \definecolor{gold3}{rgb}{0.803922,0.678431,0.000000}
  \definecolor{gold4}{rgb}{0.545098,0.458824,0.000000}
  \definecolor{goldenrod1}{rgb}{1.000000,0.756863,0.145098}
  \definecolor{goldenrod2}{rgb}{0.933333,0.705882,0.133333}
  \definecolor{goldenrod3}{rgb}{0.803922,0.607843,0.113725}
  \definecolor{goldenrod4}{rgb}{0.545098,0.411765,0.078431}
  \definecolor{DarkGoldenrod1}{rgb}{1.000000,0.725490,0.058824}
  \definecolor{DarkGoldenrod2}{rgb}{0.933333,0.678431,0.054902}
  \definecolor{DarkGoldenrod3}{rgb}{0.803922,0.584314,0.047059}
  \definecolor{DarkGoldenrod4}{rgb}{0.545098,0.396078,0.031373}
  \definecolor{RosyBrown1}{rgb}{1.000000,0.756863,0.756863}
  \definecolor{RosyBrown2}{rgb}{0.933333,0.705882,0.705882}
  \definecolor{RosyBrown3}{rgb}{0.803922,0.607843,0.607843}
  \definecolor{RosyBrown4}{rgb}{0.545098,0.411765,0.411765}
  \definecolor{IndianRed1}{rgb}{1.000000,0.415686,0.415686}
  \definecolor{IndianRed2}{rgb}{0.933333,0.388235,0.388235}
  \definecolor{IndianRed3}{rgb}{0.803922,0.333333,0.333333}
  \definecolor{IndianRed4}{rgb}{0.545098,0.227451,0.227451}
  \definecolor{sienna1}{rgb}{1.000000,0.509804,0.278431}
  \definecolor{sienna2}{rgb}{0.933333,0.474510,0.258824}
  \definecolor{sienna3}{rgb}{0.803922,0.407843,0.223529}
  \definecolor{sienna4}{rgb}{0.545098,0.278431,0.149020}
  \definecolor{burlywood1}{rgb}{1.000000,0.827451,0.607843}
  \definecolor{burlywood2}{rgb}{0.933333,0.772549,0.568627}
  \definecolor{burlywood3}{rgb}{0.803922,0.666667,0.490196}
  \definecolor{burlywood4}{rgb}{0.545098,0.450980,0.333333}
  \definecolor{wheat1}{rgb}{1.000000,0.905882,0.729412}
  \definecolor{wheat2}{rgb}{0.933333,0.847059,0.682353}
  \definecolor{wheat3}{rgb}{0.803922,0.729412,0.588235}
  \definecolor{wheat4}{rgb}{0.545098,0.494118,0.400000}
  \definecolor{tan1}{rgb}{1.000000,0.647059,0.309804}
  \definecolor{tan2}{rgb}{0.933333,0.603922,0.286275}
  \definecolor{tan3}{rgb}{0.803922,0.521569,0.247059}
  \definecolor{tan4}{rgb}{0.545098,0.352941,0.168627}
  \definecolor{chocolate1}{rgb}{1.000000,0.498039,0.141176}
  \definecolor{chocolate2}{rgb}{0.933333,0.462745,0.129412}
  \definecolor{chocolate3}{rgb}{0.803922,0.400000,0.113725}
  \definecolor{chocolate4}{rgb}{0.545098,0.270588,0.074510}
  \definecolor{firebrick1}{rgb}{1.000000,0.188235,0.188235}
  \definecolor{firebrick2}{rgb}{0.933333,0.172549,0.172549}
  \definecolor{firebrick3}{rgb}{0.803922,0.149020,0.149020}
  \definecolor{firebrick4}{rgb}{0.545098,0.101961,0.101961}
  \definecolor{brown1}{rgb}{1.000000,0.250980,0.250980}
  \definecolor{brown2}{rgb}{0.933333,0.231373,0.231373}
  \definecolor{brown3}{rgb}{0.803922,0.200000,0.200000}
  \definecolor{brown4}{rgb}{0.545098,0.137255,0.137255}
  \definecolor{salmon1}{rgb}{1.000000,0.549020,0.411765}
  \definecolor{salmon2}{rgb}{0.933333,0.509804,0.384314}
  \definecolor{salmon3}{rgb}{0.803922,0.439216,0.329412}
  \definecolor{salmon4}{rgb}{0.545098,0.298039,0.223529}
  \definecolor{LightSalmon1}{rgb}{1.000000,0.627451,0.478431}
  \definecolor{LightSalmon2}{rgb}{0.933333,0.584314,0.447059}
  \definecolor{LightSalmon3}{rgb}{0.803922,0.505882,0.384314}
  \definecolor{LightSalmon4}{rgb}{0.545098,0.341176,0.258824}
  \definecolor{orange1}{rgb}{1.000000,0.647059,0.000000}
  \definecolor{orange2}{rgb}{0.933333,0.603922,0.000000}
  \definecolor{orange3}{rgb}{0.803922,0.521569,0.000000}
  \definecolor{orange4}{rgb}{0.545098,0.352941,0.000000}
  \definecolor{DarkOrange1}{rgb}{1.000000,0.498039,0.000000}
  \definecolor{DarkOrange2}{rgb}{0.933333,0.462745,0.000000}
  \definecolor{DarkOrange3}{rgb}{0.803922,0.400000,0.000000}
  \definecolor{DarkOrange4}{rgb}{0.545098,0.270588,0.000000}
  \definecolor{coral1}{rgb}{1.000000,0.447059,0.337255}
  \definecolor{coral2}{rgb}{0.933333,0.415686,0.313726}
  \definecolor{coral3}{rgb}{0.803922,0.356863,0.270588}
  \definecolor{coral4}{rgb}{0.545098,0.243137,0.184314}
  \definecolor{tomato1}{rgb}{1.000000,0.388235,0.278431}
  \definecolor{tomato2}{rgb}{0.933333,0.360784,0.258824}
  \definecolor{tomato3}{rgb}{0.803922,0.309804,0.223529}
  \definecolor{tomato4}{rgb}{0.545098,0.211765,0.149020}
  \definecolor{OrangeRed1}{rgb}{1.000000,0.270588,0.000000}
  \definecolor{OrangeRed2}{rgb}{0.933333,0.250980,0.000000}
  \definecolor{OrangeRed3}{rgb}{0.803922,0.215686,0.000000}
  \definecolor{OrangeRed4}{rgb}{0.545098,0.145098,0.000000}
  \definecolor{red1}{rgb}{1.000000,0.000000,0.000000}
  \definecolor{red2}{rgb}{0.933333,0.000000,0.000000}
  \definecolor{red3}{rgb}{0.803922,0.000000,0.000000}
  \definecolor{red4}{rgb}{0.545098,0.000000,0.000000}
  \definecolor{DeepPink1}{rgb}{1.000000,0.078431,0.576471}
  \definecolor{DeepPink2}{rgb}{0.933333,0.070588,0.537255}
  \definecolor{DeepPink3}{rgb}{0.803922,0.062745,0.462745}
  \definecolor{DeepPink4}{rgb}{0.545098,0.039216,0.313726}
  \definecolor{HotPink1}{rgb}{1.000000,0.431373,0.705882}
  \definecolor{HotPink2}{rgb}{0.933333,0.415686,0.654902}
  \definecolor{HotPink3}{rgb}{0.803922,0.376471,0.564706}
  \definecolor{HotPink4}{rgb}{0.545098,0.227451,0.384314}
  \definecolor{pink1}{rgb}{1.000000,0.709804,0.772549}
  \definecolor{pink2}{rgb}{0.933333,0.662745,0.721569}
  \definecolor{pink3}{rgb}{0.803922,0.568627,0.619608}
  \definecolor{pink4}{rgb}{0.545098,0.388235,0.423529}
  \definecolor{LightPink1}{rgb}{1.000000,0.682353,0.725490}
  \definecolor{LightPink2}{rgb}{0.933333,0.635294,0.678431}
  \definecolor{LightPink3}{rgb}{0.803922,0.549020,0.584314}
  \definecolor{LightPink4}{rgb}{0.545098,0.372549,0.396078}
  \definecolor{PaleVioletRed1}{rgb}{1.000000,0.509804,0.670588}
  \definecolor{PaleVioletRed2}{rgb}{0.933333,0.474510,0.623529}
  \definecolor{PaleVioletRed3}{rgb}{0.803922,0.407843,0.537255}
  \definecolor{PaleVioletRed4}{rgb}{0.545098,0.278431,0.364706}
  \definecolor{maroon1}{rgb}{1.000000,0.203922,0.701961}
  \definecolor{maroon2}{rgb}{0.933333,0.188235,0.654902}
  \definecolor{maroon3}{rgb}{0.803922,0.160784,0.564706}
  \definecolor{maroon4}{rgb}{0.545098,0.109804,0.384314}
  \definecolor{VioletRed1}{rgb}{1.000000,0.243137,0.588235}
  \definecolor{VioletRed2}{rgb}{0.933333,0.227451,0.549020}
  \definecolor{VioletRed3}{rgb}{0.803922,0.196078,0.470588}
  \definecolor{VioletRed4}{rgb}{0.545098,0.133333,0.321569}
  \definecolor{magenta1}{rgb}{1.000000,0.000000,1.000000}
  \definecolor{magenta2}{rgb}{0.933333,0.000000,0.933333}
  \definecolor{magenta3}{rgb}{0.803922,0.000000,0.803922}
  \definecolor{magenta4}{rgb}{0.545098,0.000000,0.545098}
  \definecolor{orchid1}{rgb}{1.000000,0.513726,0.980392}
  \definecolor{orchid2}{rgb}{0.933333,0.478431,0.913725}
  \definecolor{orchid3}{rgb}{0.803922,0.411765,0.788235}
  \definecolor{orchid4}{rgb}{0.545098,0.278431,0.537255}
  \definecolor{plum1}{rgb}{1.000000,0.733333,1.000000}
  \definecolor{plum2}{rgb}{0.933333,0.682353,0.933333}
  \definecolor{plum3}{rgb}{0.803922,0.588235,0.803922}
  \definecolor{plum4}{rgb}{0.545098,0.400000,0.545098}
  \definecolor{MediumOrchid1}{rgb}{0.878431,0.400000,1.000000}
  \definecolor{MediumOrchid2}{rgb}{0.819608,0.372549,0.933333}
  \definecolor{MediumOrchid3}{rgb}{0.705882,0.321569,0.803922}
  \definecolor{MediumOrchid4}{rgb}{0.478431,0.215686,0.545098}
  \definecolor{DarkOrchid1}{rgb}{0.749020,0.243137,1.000000}
  \definecolor{DarkOrchid2}{rgb}{0.698039,0.227451,0.933333}
  \definecolor{DarkOrchid3}{rgb}{0.603922,0.196078,0.803922}
  \definecolor{DarkOrchid4}{rgb}{0.407843,0.133333,0.545098}
  \definecolor{purple1}{rgb}{0.607843,0.188235,1.000000}
  \definecolor{purple2}{rgb}{0.568627,0.172549,0.933333}
  \definecolor{purple3}{rgb}{0.490196,0.149020,0.803922}
  \definecolor{purple4}{rgb}{0.333333,0.101961,0.545098}
  \definecolor{MediumPurple1}{rgb}{0.670588,0.509804,1.000000}
  \definecolor{MediumPurple2}{rgb}{0.623529,0.474510,0.933333}
  \definecolor{MediumPurple3}{rgb}{0.537255,0.407843,0.803922}
  \definecolor{MediumPurple4}{rgb}{0.364706,0.278431,0.545098}
  \definecolor{thistle1}{rgb}{1.000000,0.882353,1.000000}
  \definecolor{thistle2}{rgb}{0.933333,0.823529,0.933333}
  \definecolor{thistle3}{rgb}{0.803922,0.709804,0.803922}
  \definecolor{thistle4}{rgb}{0.545098,0.482353,0.545098}
  \definecolor{gray0}{rgb}{0.000000,0.000000,0.000000}
  \definecolor{grey0}{rgb}{0.000000,0.000000,0.000000}
  \definecolor{gray1}{rgb}{0.011765,0.011765,0.011765}
  \definecolor{grey1}{rgb}{0.011765,0.011765,0.011765}
  \definecolor{gray2}{rgb}{0.019608,0.019608,0.019608}
  \definecolor{grey2}{rgb}{0.019608,0.019608,0.019608}
  \definecolor{gray3}{rgb}{0.031373,0.031373,0.031373}
  \definecolor{grey3}{rgb}{0.031373,0.031373,0.031373}
  \definecolor{gray4}{rgb}{0.039216,0.039216,0.039216}
  \definecolor{grey4}{rgb}{0.039216,0.039216,0.039216}
  \definecolor{gray5}{rgb}{0.050980,0.050980,0.050980}
  \definecolor{grey5}{rgb}{0.050980,0.050980,0.050980}
  \definecolor{gray6}{rgb}{0.058824,0.058824,0.058824}
  \definecolor{grey6}{rgb}{0.058824,0.058824,0.058824}
  \definecolor{gray7}{rgb}{0.070588,0.070588,0.070588}
  \definecolor{grey7}{rgb}{0.070588,0.070588,0.070588}
  \definecolor{gray8}{rgb}{0.078431,0.078431,0.078431}
  \definecolor{grey8}{rgb}{0.078431,0.078431,0.078431}
  \definecolor{gray9}{rgb}{0.090196,0.090196,0.090196}
  \definecolor{grey9}{rgb}{0.090196,0.090196,0.090196}
  \definecolor{gray10}{rgb}{0.101961,0.101961,0.101961}
  \definecolor{grey10}{rgb}{0.101961,0.101961,0.101961}
  \definecolor{gray11}{rgb}{0.109804,0.109804,0.109804}
  \definecolor{grey11}{rgb}{0.109804,0.109804,0.109804}
  \definecolor{gray12}{rgb}{0.121569,0.121569,0.121569}
  \definecolor{grey12}{rgb}{0.121569,0.121569,0.121569}
  \definecolor{gray13}{rgb}{0.129412,0.129412,0.129412}
  \definecolor{grey13}{rgb}{0.129412,0.129412,0.129412}
  \definecolor{gray14}{rgb}{0.141176,0.141176,0.141176}
  \definecolor{grey14}{rgb}{0.141176,0.141176,0.141176}
  \definecolor{gray15}{rgb}{0.149020,0.149020,0.149020}
  \definecolor{grey15}{rgb}{0.149020,0.149020,0.149020}
  \definecolor{gray16}{rgb}{0.160784,0.160784,0.160784}
  \definecolor{grey16}{rgb}{0.160784,0.160784,0.160784}
  \definecolor{gray17}{rgb}{0.168627,0.168627,0.168627}
  \definecolor{grey17}{rgb}{0.168627,0.168627,0.168627}
  \definecolor{gray18}{rgb}{0.180392,0.180392,0.180392}
  \definecolor{grey18}{rgb}{0.180392,0.180392,0.180392}
  \definecolor{gray19}{rgb}{0.188235,0.188235,0.188235}
  \definecolor{grey19}{rgb}{0.188235,0.188235,0.188235}
  \definecolor{gray20}{rgb}{0.200000,0.200000,0.200000}
  \definecolor{grey20}{rgb}{0.200000,0.200000,0.200000}
  \definecolor{gray21}{rgb}{0.211765,0.211765,0.211765}
  \definecolor{grey21}{rgb}{0.211765,0.211765,0.211765}
  \definecolor{gray22}{rgb}{0.219608,0.219608,0.219608}
  \definecolor{grey22}{rgb}{0.219608,0.219608,0.219608}
  \definecolor{gray23}{rgb}{0.231373,0.231373,0.231373}
  \definecolor{grey23}{rgb}{0.231373,0.231373,0.231373}
  \definecolor{gray24}{rgb}{0.239216,0.239216,0.239216}
  \definecolor{grey24}{rgb}{0.239216,0.239216,0.239216}
  \definecolor{gray25}{rgb}{0.250980,0.250980,0.250980}
  \definecolor{grey25}{rgb}{0.250980,0.250980,0.250980}
  \definecolor{gray26}{rgb}{0.258824,0.258824,0.258824}
  \definecolor{grey26}{rgb}{0.258824,0.258824,0.258824}
  \definecolor{gray27}{rgb}{0.270588,0.270588,0.270588}
  \definecolor{grey27}{rgb}{0.270588,0.270588,0.270588}
  \definecolor{gray28}{rgb}{0.278431,0.278431,0.278431}
  \definecolor{grey28}{rgb}{0.278431,0.278431,0.278431}
  \definecolor{gray29}{rgb}{0.290196,0.290196,0.290196}
  \definecolor{grey29}{rgb}{0.290196,0.290196,0.290196}
  \definecolor{gray30}{rgb}{0.301961,0.301961,0.301961}
  \definecolor{grey30}{rgb}{0.301961,0.301961,0.301961}
  \definecolor{gray31}{rgb}{0.309804,0.309804,0.309804}
  \definecolor{grey31}{rgb}{0.309804,0.309804,0.309804}
  \definecolor{gray32}{rgb}{0.321569,0.321569,0.321569}
  \definecolor{grey32}{rgb}{0.321569,0.321569,0.321569}
  \definecolor{gray33}{rgb}{0.329412,0.329412,0.329412}
  \definecolor{grey33}{rgb}{0.329412,0.329412,0.329412}
  \definecolor{gray34}{rgb}{0.341176,0.341176,0.341176}
  \definecolor{grey34}{rgb}{0.341176,0.341176,0.341176}
  \definecolor{gray35}{rgb}{0.349020,0.349020,0.349020}
  \definecolor{grey35}{rgb}{0.349020,0.349020,0.349020}
  \definecolor{gray36}{rgb}{0.360784,0.360784,0.360784}
  \definecolor{grey36}{rgb}{0.360784,0.360784,0.360784}
  \definecolor{gray37}{rgb}{0.368627,0.368627,0.368627}
  \definecolor{grey37}{rgb}{0.368627,0.368627,0.368627}
  \definecolor{gray38}{rgb}{0.380392,0.380392,0.380392}
  \definecolor{grey38}{rgb}{0.380392,0.380392,0.380392}
  \definecolor{gray39}{rgb}{0.388235,0.388235,0.388235}
  \definecolor{grey39}{rgb}{0.388235,0.388235,0.388235}
  \definecolor{gray40}{rgb}{0.400000,0.400000,0.400000}
  \definecolor{grey40}{rgb}{0.400000,0.400000,0.400000}
  \definecolor{gray41}{rgb}{0.411765,0.411765,0.411765}
  \definecolor{grey41}{rgb}{0.411765,0.411765,0.411765}
  \definecolor{gray42}{rgb}{0.419608,0.419608,0.419608}
  \definecolor{grey42}{rgb}{0.419608,0.419608,0.419608}
  \definecolor{gray43}{rgb}{0.431373,0.431373,0.431373}
  \definecolor{grey43}{rgb}{0.431373,0.431373,0.431373}
  \definecolor{gray44}{rgb}{0.439216,0.439216,0.439216}
  \definecolor{grey44}{rgb}{0.439216,0.439216,0.439216}
  \definecolor{gray45}{rgb}{0.450980,0.450980,0.450980}
  \definecolor{grey45}{rgb}{0.450980,0.450980,0.450980}
  \definecolor{gray46}{rgb}{0.458824,0.458824,0.458824}
  \definecolor{grey46}{rgb}{0.458824,0.458824,0.458824}
  \definecolor{gray47}{rgb}{0.470588,0.470588,0.470588}
  \definecolor{grey47}{rgb}{0.470588,0.470588,0.470588}
  \definecolor{gray48}{rgb}{0.478431,0.478431,0.478431}
  \definecolor{grey48}{rgb}{0.478431,0.478431,0.478431}
  \definecolor{gray49}{rgb}{0.490196,0.490196,0.490196}
  \definecolor{grey49}{rgb}{0.490196,0.490196,0.490196}
  \definecolor{gray50}{rgb}{0.498039,0.498039,0.498039}
  \definecolor{grey50}{rgb}{0.498039,0.498039,0.498039}
  \definecolor{gray51}{rgb}{0.509804,0.509804,0.509804}
  \definecolor{grey51}{rgb}{0.509804,0.509804,0.509804}
  \definecolor{gray52}{rgb}{0.521569,0.521569,0.521569}
  \definecolor{grey52}{rgb}{0.521569,0.521569,0.521569}
  \definecolor{gray53}{rgb}{0.529412,0.529412,0.529412}
  \definecolor{grey53}{rgb}{0.529412,0.529412,0.529412}
  \definecolor{gray54}{rgb}{0.541176,0.541176,0.541176}
  \definecolor{grey54}{rgb}{0.541176,0.541176,0.541176}
  \definecolor{gray55}{rgb}{0.549020,0.549020,0.549020}
  \definecolor{grey55}{rgb}{0.549020,0.549020,0.549020}
  \definecolor{gray56}{rgb}{0.560784,0.560784,0.560784}
  \definecolor{grey56}{rgb}{0.560784,0.560784,0.560784}
  \definecolor{gray57}{rgb}{0.568627,0.568627,0.568627}
  \definecolor{grey57}{rgb}{0.568627,0.568627,0.568627}
  \definecolor{gray58}{rgb}{0.580392,0.580392,0.580392}
  \definecolor{grey58}{rgb}{0.580392,0.580392,0.580392}
  \definecolor{gray59}{rgb}{0.588235,0.588235,0.588235}
  \definecolor{grey59}{rgb}{0.588235,0.588235,0.588235}
  \definecolor{gray60}{rgb}{0.600000,0.600000,0.600000}
  \definecolor{grey60}{rgb}{0.600000,0.600000,0.600000}
  \definecolor{gray61}{rgb}{0.611765,0.611765,0.611765}
  \definecolor{grey61}{rgb}{0.611765,0.611765,0.611765}
  \definecolor{gray62}{rgb}{0.619608,0.619608,0.619608}
  \definecolor{grey62}{rgb}{0.619608,0.619608,0.619608}
  \definecolor{gray63}{rgb}{0.631373,0.631373,0.631373}
  \definecolor{grey63}{rgb}{0.631373,0.631373,0.631373}
  \definecolor{gray64}{rgb}{0.639216,0.639216,0.639216}
  \definecolor{grey64}{rgb}{0.639216,0.639216,0.639216}
  \definecolor{gray65}{rgb}{0.650980,0.650980,0.650980}
  \definecolor{grey65}{rgb}{0.650980,0.650980,0.650980}
  \definecolor{gray66}{rgb}{0.658824,0.658824,0.658824}
  \definecolor{grey66}{rgb}{0.658824,0.658824,0.658824}
  \definecolor{gray67}{rgb}{0.670588,0.670588,0.670588}
  \definecolor{grey67}{rgb}{0.670588,0.670588,0.670588}
  \definecolor{gray68}{rgb}{0.678431,0.678431,0.678431}
  \definecolor{grey68}{rgb}{0.678431,0.678431,0.678431}
  \definecolor{gray69}{rgb}{0.690196,0.690196,0.690196}
  \definecolor{grey69}{rgb}{0.690196,0.690196,0.690196}
  \definecolor{gray70}{rgb}{0.701961,0.701961,0.701961}
  \definecolor{grey70}{rgb}{0.701961,0.701961,0.701961}
  \definecolor{gray71}{rgb}{0.709804,0.709804,0.709804}
  \definecolor{grey71}{rgb}{0.709804,0.709804,0.709804}
  \definecolor{gray72}{rgb}{0.721569,0.721569,0.721569}
  \definecolor{grey72}{rgb}{0.721569,0.721569,0.721569}
  \definecolor{gray73}{rgb}{0.729412,0.729412,0.729412}
  \definecolor{grey73}{rgb}{0.729412,0.729412,0.729412}
  \definecolor{gray74}{rgb}{0.741176,0.741176,0.741176}
  \definecolor{grey74}{rgb}{0.741176,0.741176,0.741176}
  \definecolor{gray75}{rgb}{0.749020,0.749020,0.749020}
  \definecolor{grey75}{rgb}{0.749020,0.749020,0.749020}
  \definecolor{gray76}{rgb}{0.760784,0.760784,0.760784}
  \definecolor{grey76}{rgb}{0.760784,0.760784,0.760784}
  \definecolor{gray77}{rgb}{0.768627,0.768627,0.768627}
  \definecolor{grey77}{rgb}{0.768627,0.768627,0.768627}
  \definecolor{gray78}{rgb}{0.780392,0.780392,0.780392}
  \definecolor{grey78}{rgb}{0.780392,0.780392,0.780392}
  \definecolor{gray79}{rgb}{0.788235,0.788235,0.788235}
  \definecolor{grey79}{rgb}{0.788235,0.788235,0.788235}
  \definecolor{gray80}{rgb}{0.800000,0.800000,0.800000}
  \definecolor{grey80}{rgb}{0.800000,0.800000,0.800000}
  \definecolor{gray81}{rgb}{0.811765,0.811765,0.811765}
  \definecolor{grey81}{rgb}{0.811765,0.811765,0.811765}
  \definecolor{gray82}{rgb}{0.819608,0.819608,0.819608}
  \definecolor{grey82}{rgb}{0.819608,0.819608,0.819608}
  \definecolor{gray83}{rgb}{0.831373,0.831373,0.831373}
  \definecolor{grey83}{rgb}{0.831373,0.831373,0.831373}
  \definecolor{gray84}{rgb}{0.839216,0.839216,0.839216}
  \definecolor{grey84}{rgb}{0.839216,0.839216,0.839216}
  \definecolor{gray85}{rgb}{0.850980,0.850980,0.850980}
  \definecolor{grey85}{rgb}{0.850980,0.850980,0.850980}
  \definecolor{gray86}{rgb}{0.858824,0.858824,0.858824}
  \definecolor{grey86}{rgb}{0.858824,0.858824,0.858824}
  \definecolor{gray87}{rgb}{0.870588,0.870588,0.870588}
  \definecolor{grey87}{rgb}{0.870588,0.870588,0.870588}
  \definecolor{gray88}{rgb}{0.878431,0.878431,0.878431}
  \definecolor{grey88}{rgb}{0.878431,0.878431,0.878431}
  \definecolor{gray89}{rgb}{0.890196,0.890196,0.890196}
  \definecolor{grey89}{rgb}{0.890196,0.890196,0.890196}
  \definecolor{gray90}{rgb}{0.898039,0.898039,0.898039}
  \definecolor{grey90}{rgb}{0.898039,0.898039,0.898039}
  \definecolor{gray91}{rgb}{0.909804,0.909804,0.909804}
  \definecolor{grey91}{rgb}{0.909804,0.909804,0.909804}
  \definecolor{gray92}{rgb}{0.921569,0.921569,0.921569}
  \definecolor{grey92}{rgb}{0.921569,0.921569,0.921569}
  \definecolor{gray93}{rgb}{0.929412,0.929412,0.929412}
  \definecolor{grey93}{rgb}{0.929412,0.929412,0.929412}
  \definecolor{gray94}{rgb}{0.941176,0.941176,0.941176}
  \definecolor{grey94}{rgb}{0.941176,0.941176,0.941176}
  \definecolor{gray95}{rgb}{0.949020,0.949020,0.949020}
  \definecolor{grey95}{rgb}{0.949020,0.949020,0.949020}
  \definecolor{gray96}{rgb}{0.960784,0.960784,0.960784}
  \definecolor{grey96}{rgb}{0.960784,0.960784,0.960784}
  \definecolor{gray97}{rgb}{0.968627,0.968627,0.968627}
  \definecolor{grey97}{rgb}{0.968627,0.968627,0.968627}
  \definecolor{gray98}{rgb}{0.980392,0.980392,0.980392}
  \definecolor{grey98}{rgb}{0.980392,0.980392,0.980392}
  \definecolor{gray99}{rgb}{0.988235,0.988235,0.988235}
  \definecolor{grey99}{rgb}{0.988235,0.988235,0.988235}
  \definecolor{gray100}{rgb}{1.000000,1.000000,1.000000}
  \definecolor{grey100}{rgb}{1.000000,1.000000,1.000000}
  \definecolor{dark grey}{rgb}{0.662745,0.662745,0.662745}
  \definecolor{DarkGrey}{rgb}{0.662745,0.662745,0.662745}
  \definecolor{dark gray}{rgb}{0.662745,0.662745,0.662745}
  \definecolor{DarkGray}{rgb}{0.662745,0.662745,0.662745}
  \definecolor{dark blue}{rgb}{0.000000,0.000000,0.545098}
  \definecolor{DarkBlue}{rgb}{0.000000,0.000000,0.545098}
  \definecolor{dark cyan}{rgb}{0.000000,0.545098,0.545098}
  \definecolor{DarkCyan}{rgb}{0.000000,0.545098,0.545098}
  \definecolor{dark magenta}{rgb}{0.545098,0.000000,0.545098}
  \definecolor{DarkMagenta}{rgb}{0.545098,0.000000,0.545098}
  \definecolor{dark red}{rgb}{0.545098,0.000000,0.000000}
  \definecolor{DarkRed}{rgb}{0.545098,0.000000,0.000000}
  \definecolor{light green}{rgb}{0.564706,0.933333,0.564706}
  \definecolor{LightGreen}{rgb}{0.564706,0.933333,0.564706}

\begin{document}

\preprint{}

\title{Isospin transport in $^{84}$Kr + $^{112,124}$Sn
collisions at Fermi energies}

\author{S.~Barlini}
\affiliation{Sezione INFN di Firenze, Via G. Sansone 1, 
     I-50019 Sesto Fiorentino, Italy}
\affiliation{Dipartimento di Fisica, Univ. di Firenze, 
     Via G. Sansone 1, I-50019 Sesto Fiorentino, Italy}

\author{S.~Piantelli}
\affiliation{Sezione INFN di Firenze, Via G. Sansone 1, 
     I-50019 Sesto Fiorentino, Italy}

\author{G.~Casini}
\affiliation{Sezione INFN di Firenze, Via G. Sansone 1, 
     I-50019 Sesto Fiorentino, Italy}

\author{P.R.~Maurenzig}
\affiliation{Sezione INFN di Firenze, Via G. Sansone 1, 
     I-50019 Sesto Fiorentino, Italy}
\affiliation{Dipartimento di Fisica, Univ. di Firenze, 
     Via G. Sansone 1, I-50019 Sesto Fiorentino, Italy}

\author{A.~Olmi}
\thanks{corresponding author} \email[e-mail:]{olmi@fi.infn.it}
\affiliation{Sezione INFN di Firenze, Via G. Sansone 1, 
     I-50019 Sesto Fiorentino, Italy}

\author{M.~Bini}
\affiliation{Sezione INFN di Firenze, Via G. Sansone 1, 
     I-50019 Sesto Fiorentino, Italy}
\affiliation{Dipartimento di Fisica, Univ. di Firenze, 
     Via G. Sansone 1, I-50019 Sesto Fiorentino, Italy}

\author{S.~Carboni}
\affiliation{Sezione INFN di Firenze, Via G. Sansone 1, 
     I-50019 Sesto Fiorentino, Italy}
\affiliation{Dipartimento di Fisica, Univ. di Firenze, 
     Via G. Sansone 1, I-50019 Sesto Fiorentino, Italy}

\author{G.~Pasquali}
\affiliation{Sezione INFN di Firenze, Via G. Sansone 1, 
     I-50019 Sesto Fiorentino, Italy}
\affiliation{Dipartimento di Fisica, Univ. di Firenze, 
     Via G. Sansone 1, I-50019 Sesto Fiorentino, Italy}

\author{G.~Poggi}
\affiliation{Sezione INFN di Firenze, Via G. Sansone 1, 
     I-50019 Sesto Fiorentino, Italy}
\affiliation{Dipartimento di Fisica, Univ. di Firenze, 
     Via G. Sansone 1, I-50019 Sesto Fiorentino, Italy}

\author{A.A.~Stefanini}
\affiliation{Sezione INFN di Firenze, Via G. Sansone 1, 
     I-50019 Sesto Fiorentino, Italy}
\affiliation{Dipartimento di Fisica, Univ. di Firenze, 
     Via G. Sansone 1, I-50019 Sesto Fiorentino, Italy}

\author{R.~Bougault}
\affiliation{LPC, IN2P3-CNRS, ENSICAEN et Universit\'{e} de Caen, 
 F-14050 Caen-Cedex, France}

\author{E.~Bonnet}
\affiliation{GANIL, CEA/DSM-CNRS/IN2P3, B.P. 5027, F-14076 Caen cedex, France}

\author{B.~Borderie}
\affiliation{Institut de Physique Nucl\'{e}aire, CNRS/IN2P3, 
      Universit\'{e} Paris-Sud 11, F-91406 Orsay cedex, France}

\author{A.~Chbihi}
\affiliation{GANIL, CEA/DSM-CNRS/IN2P3, B.P. 5027, F-14076 Caen cedex, France}

\author{J.D.~Frankland}
\affiliation{GANIL, CEA/DSM-CNRS/IN2P3, B.P. 5027, F-14076 Caen cedex, France}

\author{D.~Gruyer}
\affiliation{GANIL, CEA/DSM-CNRS/IN2P3, B.P. 5027, F-14076 Caen cedex, France}

\author{O.~Lopez}
\affiliation{LPC, IN2P3-CNRS, ENSICAEN et Universit\'{e} de Caen, 
 F-14050 Caen-Cedex, France}

\author{N.~Le~Neindre}
\affiliation{LPC, IN2P3-CNRS, ENSICAEN et Universit\'{e} de Caen, 
 F-14050 Caen-Cedex, France}

\author{M.~P\^{a}rlog}
\affiliation{LPC, IN2P3-CNRS, ENSICAEN et Universit\'{e} de Caen, 
 F-14050 Caen-Cedex, France}
\affiliation{Horia Hulubei, National Institute of Physics and 
    Nuclear Engineering, RO-077125 Bucharest-M\u{a}gurele, Romania}

\author{M.F.~Rivet}
\affiliation{Institut de Physique Nucl\'{e}aire, CNRS/IN2P3, 
      Universit\'{e} Paris-Sud 11, F-91406 Orsay cedex, France}

\author{E.~Vient}
\affiliation{LPC, IN2P3-CNRS, ENSICAEN et Universit\'{e} de Caen, 
 F-14050 Caen-Cedex, France}

\author{E.~Rosato}
\affiliation{Dipartimento di Scienze Fisiche e Sezione INFN, 
    Universit\'{a} di Napoli ``Federico II'', I 80126 Napoli, Italy}

\author{G.~Spadaccini}
\affiliation{Dipartimento di Scienze Fisiche e Sezione INFN, 
    Universit\'{a} di Napoli ``Federico II'', I 80126 Napoli, Italy}

\author{M.~Vigilante}
\affiliation{Dipartimento di Scienze Fisiche e Sezione INFN, 
    Universit\'{a} di Napoli ``Federico II'', I 80126 Napoli, Italy}

\author{M.~Bruno}
\affiliation{INFN e Universit\'{a} di Bologna, 40126 Bologna, Italy}

\author{T.~Marchi}
\affiliation{INFN LNL Legnaro, viale dell'Universit\'{a} 2, 
      35020 Legnaro (Padova) Italy}

\author{L.~Morelli}
\affiliation{INFN e Universit\'{a} di Bologna, 40126 Bologna, Italy}

\author{M.~Cinausero}
\affiliation{INFN LNL Legnaro, viale dell'Universit\'{a} 2, 
      35020 Legnaro (Padova) Italy}

\author{M.~Degerlier}
\affiliation{INFN LNL Legnaro, viale dell'Universit\'{a} 2, 
      35020 Legnaro (Padova) Italy}

\author{F.~Gramegna}
\affiliation{INFN LNL Legnaro, viale dell'Universit\'{a} 2, 
      35020 Legnaro (Padova) Italy}

\author{T.~Kozik}
\affiliation{Jagiellonian University, Institute of Nuclear Physics 
     IFJ-PAN, PL-31342 Krak\'{o}w, Poland}

\author{T.~Twar\'{o}g}
\affiliation{Jagiellonian University, Institute of Nuclear Physics 
     IFJ-PAN, PL-31342 Krak\'{o}w, Poland}

\author{R.~Alba}
\affiliation{INFN LNS, via S.Sofia 62, 95125 Catania, Italy}

\author{C.~Maiolino}
\affiliation{INFN LNS, via S.Sofia 62, 95125 Catania, Italy}

\author{D.~Santonocito}
\affiliation{INFN LNS, via S.Sofia 62, 95125 Catania, Italy}

\collaboration{FAZIA Collaboration}
\noaffiliation

\date{\today}

\begin{abstract}
Isotopically resolved fragments with Z$\alt$20 have been studied 
with high resolution telescopes in a test run for the FAZIA collaboration.
The fragments were produced by the collision of a $^{84}$Kr beam at 35 
MeV/nucleon with a n-rich ($^{124}$Sn) and a n-poor ($^{112}$Sn) target.
The fragments, detected close to the grazing angle, are mainly emitted 
from the phase-space region of the projectile.
The fragment isotopic content clearly depends on the n-richness of the target 
and it is a direct evidence of isospin diffusion between projectile and target.
The observed enhanced neutron richness of light fragments emitted 
from the phase-space region close to the center of mass of the system
can be interpreted as an effect of isospin drift in the diluted neck region.
\end{abstract}

\maketitle
\section{INTRODUCTION}
\label{sec:Introduction}

The production of many fragments with different sizes is one of the
main features of heavy-ion reactions at bombarding energies
higher than 15-20 MeV/u. 
The mechanisms governing their production have been extensively 
investigated in the past.
When the primary fragments produced in the interaction are sufficiently 
excited, their detection occurs after a de-excitation phase that may 
strongly alter their original identity.
Various de-excitation processes are indeed possible and they depend both 
on the initial conditions and on the internal structure of 
the nuclei involved in the de-excitation path. 

In recent years many experimental and theoretical 
(see \cite{WCI,shetty07,galich09e,baoanli08,ditoro10} and references therein)
efforts have been devoted to the investigation of the 
neutron-to-proton ratio N/Z (often called isospin) degree of freedom 
and to unravelling its influence on the reaction dynamics and
on the subsequent decay processes.
This was obtained either by using reaction partners with different isospin 
content or by comparing data from reactions involving 
different isotopic combinations of the projectile and/or of the target 
\cite{rami00,vese00,geraci04,souliotis04,tsang04,tsang09,sun10,lomba11,defili12}.
From an experimental point of view, this kind of investigation requires
detectors capable of good isotopic identification of the reaction products
on an extended Z range.

The study of the isospin content of the emitted fragments and light particles,
possibly complemented by a characterization of their emitting source 
\cite{fiasco06,mcin10,defili12},
gives clues on different processes of isospin transport.
One, called isospin ``diffusion'', is related to the isospin asymmetry of a
system in which projectile and target have different N/Z values
\cite{rami00,tsang04,baran05,chen05,baoanli08,liu07,tsang09,sun10,ditoro10}; 
the other, called isospin ``drift'' (or ``migration''), 
is related to the density gradient
which is expected to exist in the ``neck'' region, even between two
identical nuclei 
\cite{baran05,lionti05,theri06,fiasco06,shetty07,napolitani2010}. 
In both cases the experimental observables associated with the isospin 
content of the reaction products can be used to extract information on the 
symmetry energy term of the nuclear equation of state, via comparison 
with theoretical models 
\cite{WCI,isospin,souliotis03,souliotis04,tsang04,baran05,chen05,shetty07,liu07,baoanli08,colonna08,chen09,galich09t,tsang09,ditoro10,defili12}.

Many experiments found evidences of isospin transport in dissipative
collisions at Fermi energies \cite{vese00,tsang04,theri06,liu07,tsang09,galich09e,mcin10,sun10,defili12}. 
In this paper we show some results obtained by bombarding with a $^{84}$Kr beam
at 35 MeV/u two targets with different isospin: $^{112}$Sn and $^{124}$Sn.
In the following we will often use ``n-poor'' and ``n-rich system'' 
to refer to the collision of the Kr beam with these two different targets.
Although the 
light complex fragments detected in our experiment
originate mainly from the 
quasi-projectile source, their isospin content shows a clear dependence 
on the target isotope.

\section{THE EXPERIMENT }        \label{sec:experimental_apparatus}

The data presented in this paper have been collected 
by the FAZIA collaboration \cite{Fazia}
at the Superconducting Cyclotron of the Laboratori Nazionali del Sud 
(LNS) of INFN, in Catania,
during a recent test experiment \cite{fr12}.
A pulsed beam ($\delta$t $\approx$ 1 ns FWHM) of $^{84}$Kr at 35 MeV/nucleon
impinged on isotopically enriched targets of $^{112}$Sn (415 $\mu$g/cm$^2$) 
and $^{124}$Sn (600 $\mu$g/cm$^2$).
The N/Z of the beam was 1.33, intermediate between that of the two 
targets of $^{112}$Sn (N/Z=1.24) and $^{124}$Sn (N/Z=1.48).
In the past these systems (and other similar Kr or Sn induced reactions, in 
direct or reverse kinematics) have been the
subject of extensive investigations at comparable bombarding energies 
by other groups \cite{vese00,souliotis03,souliotis04,tsang04,liu07,tsang09,sun10,mcin10,defili12}, so that they represent a good 
benchmark for a test experiment.

Here we analyze the data of a 
three-element telescope (Si1-Si2-CsI(Tl)) 
located at an angle of 5.4$^\circ$ and at 100 cm distance from the target.
The silicon detectors (manufactured by FBK \cite{FBK}) were ion-implanted
of the neutron transmutation doped (n-TD) type, with bulk resistivity values 
in the range 3000--4000 $\Omega$cm and good doping uniformity 
(of the order of 3\% FWHM \cite{bard09a}).
The silicon layers were obtained from ``random'' cut wafers (about 7$^\circ$ 
off the $<$100$>$ axis) to minimize channeling effects \cite{bard09b}.
They were in transmission mounting, with dead layers on both sides 
of $\sim$500--800 nm and had an active area of 20$\times$20 mm$^2$.
The thickness of Si1 and Si2 was 305 $\mu$m and 510 $\mu$m, respectively,
with a measured non-uniformity of the order of 1 $\mu$m.
The CsI(Tl) crystal (manufactured by Amcrys \cite{amcrys}) 
was 10 cm thick, with an excellent doping uniformity 
(of the order of 5\%), 
and it was read out by a photodiode.
The telescope was equipped with custom-built high-quality electronics.
More details on the characteristics of the setup and on the obtained 
performances are given elsewhere 
\cite{fr12,bard09a,bard09b,bard11,Car12,pasquali12,rd12}.
Here we briefly remind that the charge and current signals produced in 
low-noise preamplifiers \cite{paci}, mounted in vacuum next to the detectors, 
are sampled by fast digital boards purposely built by the FAZIA group.
For each detector, the sampled signals are then stored for off-line analysis.
Energy information from the two silicon detectors was obtained by means of 
trapezoidal shaping of the digitized signals (see \cite{Car12} for details).
For energy calibration, the socalled ``punch-through energies'' \cite{bra11} 
of light identified ions were used, as described also in \cite{Car12}.

In this work we concentrate on identified fragments (Z$\geq$3) 
that are stopped in the second silicon layer or in the CsI(Tl).
The kinetic energy of fragments stopped in Si2 is the sum of the two 
silicon energies E$_\mathrm{\,sum}$=E$_1$+E$_2$.
When reaching the CsI(Tl) crystal, the full kinetic energy E is
estimated from E$_\mathrm{\,sum}$ (which is now the energy-loss over the 
known total thickness of the two silicon detectors) 
with the help of range-energy tables \cite{Northcliffe,Hubert,INDRA}, 
whose proper use requires the knowledge of Z and A of the ion. 

\begin{figure}[t!]
\begin{center}
\includegraphics[width=8.4cm] {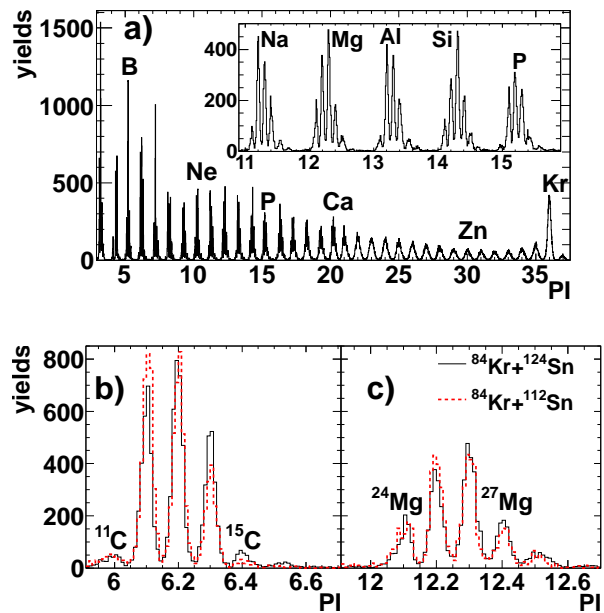}      
\end{center}
\caption{(Color online) 
  (a) Particle Identification (PI) spectrum for fragments passing
  the first silicon detector and stopped in the second one 
  or in the CsI(Tl) crystal, for the reaction $^{84}$Kr+$^{124}$Sn; 
  the inset is an expansion of the region Z=11--15.
  (b) PI spectra for Carbon isotopes in the reactions
  $^{84}$Kr+$^{124}$Sn (black histogram) and $^{84}$Kr+$^{112}$Sn (dashed red);
  the spectra are normalized to the same total number of counts of C.
  (c) Same as (b), but for Mg isotopes.} 
\label{fig_pi}
\end{figure}

The particle identification is given by the ridges in the correlations
E$_1$--E$_2$ between the energies of the two silicon layers,
or E$_\mathrm{\,sum}$--LO between the silicon energy 
and the Light-Output of the CsI(Tl).
The linearization of the ridges gives the socalled 
Particle Identification (PI).
The high quality of the detectors and of the dedicated electronics
allows isotopic resolution up to Z$\approx$20 
(close to the limit reported in \cite{Car12}),
as shown in Fig. \ref{fig_pi}(a) by the PI spectrum for the n-rich target
(the inset is a zoom of the region between Z=11 and Z=16).
Figures \ref{fig_pi}(b) and (c) are the PI spectra for C and Mg isotopes,
respectively.
The black solid histograms correspond to the n-rich target and the red
dashed ones to the n-poor target.
For each element, the two histograms are normalized to the same number 
of counts.
One sees at first glance that the isotopic composition is different
in the two reactions.
For each element, mass values are assigned to the PI peaks by comparing the 
isotopic ridges in the already mentioned correlations with the 
theoretical lines calculated from energy-loss tables.

\section{RESULTS}

The telescope spanned the angular range from about 4.8$^\circ$ to 6$^\circ$, 
just beyond the grazing angles of the two reactions (estimated to be about 
4.1$^\circ$ and 4.0$^\circ$ for the n-poor and n-rich system, respectively), 
therefore its position was well suited for 
a good sampling of a large variety of fragments, mainly originating from the 
quasi-projectile (QP) phase-space.  
We want to stress that the beam and the setup are the same, 
the kinematics is very similar and the only relevant difference between 
the two systems is the neutron number of the target nucleus. 
Since we are mainly dealing with fragments originating from the QP
phase-space, any substantial difference between the two sets of data has to 
be attributed to a transport of isospin between projectile and target.

From the large number of experiments performed in many years of investigation 
of heavy-ion collisions in the Fermi energy regime, we now know that:
a) we mainly deal with binary dissipative collisions producing excited
 quasi-projectiles (QP) and quasi-targets (QT);
b) their decay is dominated by evaporation, in competition with
 fission-like processes, especially for massive nuclei or large excitations;
c) the most central collisions involve fusion-like phenomena, with the 
 formation of a big transient system, which may then undergo a
 multifragmentation decay;
d) non-equilibrium phenomena are present,
 consisting in the rapid emission of light reaction products
 (neck emissions \cite{pian02,baran04}), or in the occurrence
 of fission-like processes retaining some memory of the preceding dynamics
 (fast oriented fission \cite{casi93,stefa95,mcin10,wilcz10}).

\begin{figure}[t!]
\begin{center}
\includegraphics[width=8cm,bb=0 0 567 360] {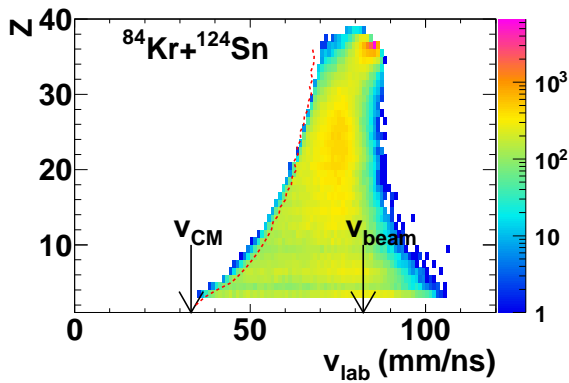}    
\end{center}
\caption{(Color online) Charge vs. laboratory-velocity for Z$\geq$3 fragments 
  passing the first silicon detector in the reaction $^{84}$Kr+$^{124}$Sn.
  Arrows indicate the center-of-mass and beam velocities.
  The red dashed line is the expected threshold due to the first 
  silicon detector.}
\label{zetavel}
\end{figure}

The origin of the detected reaction products is often deduced
from the correlation charge vs. laboratory velocity
(see, e.g., \cite{pag01,galich09e,defili09,mcin10}). 
An example is shown in Fig. \ref{zetavel} for the reaction $^{84}$Kr+$^{124}$Sn
(a similar plot is obtained also for $^{84}$Kr+$^{112}$Sn).
The laboratory velocity is deduced from the measured energy, 
using the identified mass (up to $Z\sim 20$) 
or the mass estimated from
 the Evaporation Attractor Line (EAL) \cite{char98} for heavier elements.

The dashed line indicates the estimated Z-dependent threshold due to the 
requirement of passing through the first silicon detector.
The arrows, corresponding to the velocities of the center of mass
and of the beam (33.2 and 82.2 mm/ns, respectively), 
indicate that practically all measured fragments are 
forward-emitted in the center-of-mass system and that the velocities 
of the heavier ones are not too different from that of the projectile.
Therefore one can infer that the fragments originate indeed from the QP, 
with almost no contamination from the QT (in the analysis we reject very 
light fragments with v$_\mathrm{lab}<$40 mm/ns), and that there could be 
-at most- some contribution from the ``neck region'' (i.e. the phase space 
region corresponding to the contact zone of the colliding nuclei).

\begin{figure}[t!]
\begin{center}
\includegraphics[width=7.5cm] {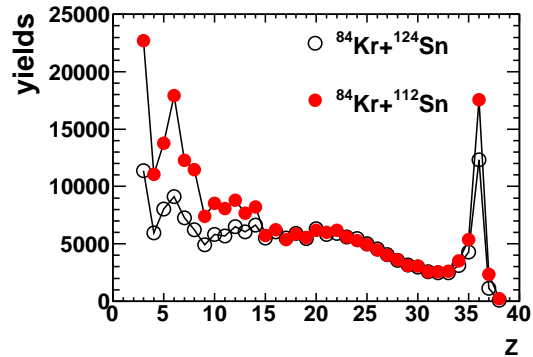}      
\end{center}
\caption{(Color online) Inclusive charge distribution of fragments (Z$\geq3$)
  produced in the reactions of $^{84}$Kr on $^{124}$Sn (black open dots) and
  $^{112}$Sn (red solid dots) at 35 MeV/u. 
  The distributions are normalized in the region 18$\leq$Z$\leq$28 (see text).}
\label{chargedis}
\end{figure}

\begin{figure*}[t!]
\begin{center}
\includegraphics[width=14.5cm] {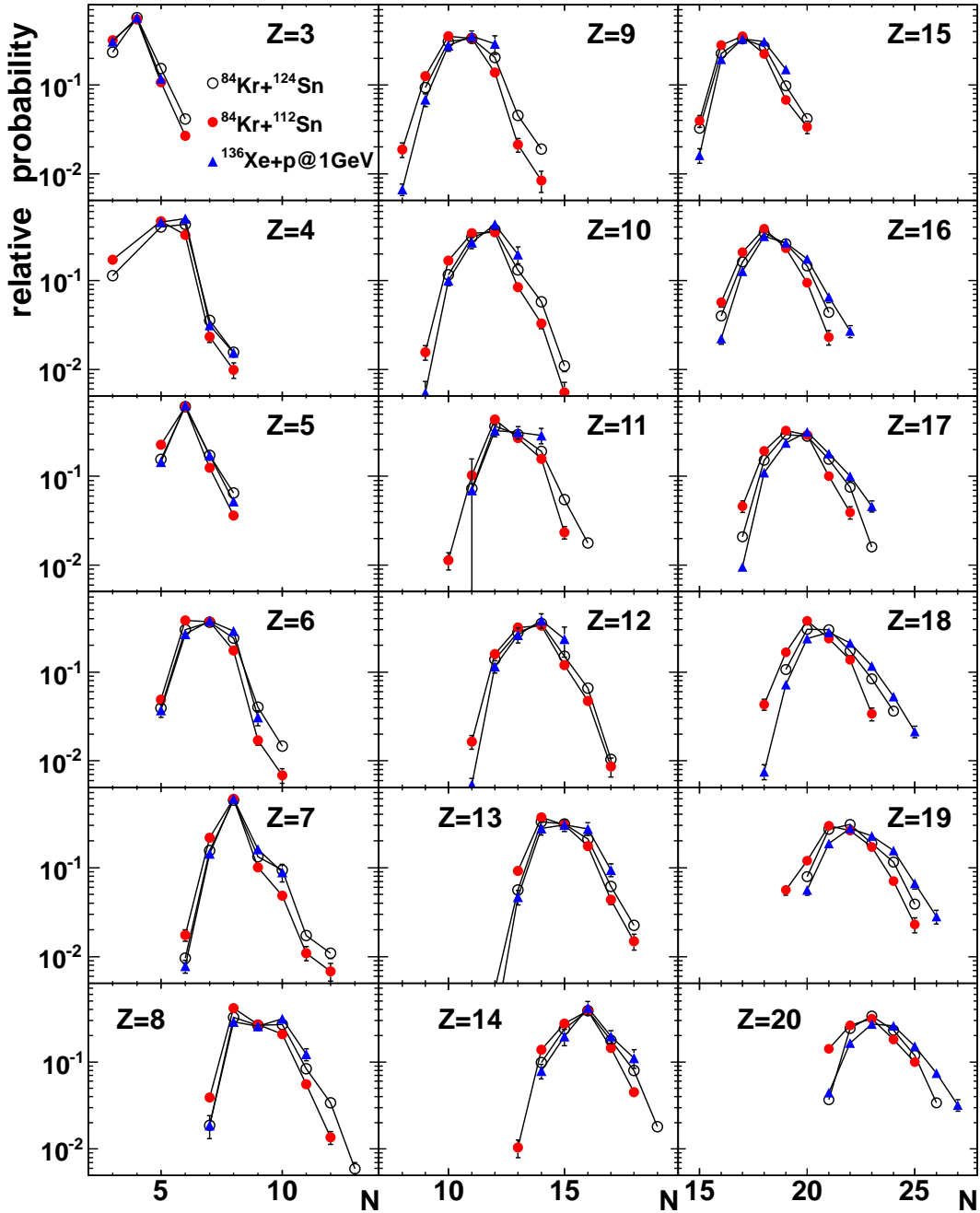}       
\end{center}
\caption{(Color online) Relative probability of isotopic population for
  elements with Z=3--20,
  obtained by normalizing each isotopic distribution to unity.
  The data are from the reactions $^{84}$Kr+$^{124}$Sn (open black dots)
  and $^{84}$Kr+$^{112}$Sn (solid red dots) at 35 MeV/u, and from
  $^{136}$Xe+H (blue triangles) at 1 GeV/u \cite{Nap2007}.
  Bars represent statistical errors.}
\label{fig_perc}
\end{figure*}

From the quasi-elastic peak (near Z=36 and v=82.2 mm/ns),
an evident ridge (only marginally affected by the threshold)
develops towards lower velocities and lighter fragments. 
This is a characteristic feature of binary dissipative 
collisions \cite{huiz84,mcin10}: 
with decreasing velocity, the QP excitation increases, 
so that it is detected as a lighter QP remnant
\cite{piant08,theri06} after a long decay chain.
Indeed statistical calculations with the code GEMINI \cite{gemini}
show that an excited $^{84}$Kr nucleus, with a typical excitation energy of 
300 MeV and spin 30 $\hbar$, ends up in a bell-shaped distribution
centered at Z$\sim$28, with a tail extending down to Z$\sim$20.
Of course, because of possibly early dynamical emissions, the 
evaporating QP can be somewhat lighter than the projectile; 
as an example, in $^{64}$Zn+$^{64}$Zn at 45 MeV/nucleon \cite{theri06}
the reconstructed primary QP charges were $\sim$20$\%$ smaller 
than Z=30, already for moderate dissipations. 
In Fig. \ref{zetavel} the most probable velocity of the ridge saturates at
$\sim$75 mm/ns in correspondence with 
the broad charge distribution visible around Z=15--25.
Assuming a binary kinematics, as it was done in \cite{souliotis04}, one
would estimate a dissipation of about 500 MeV and an average excitation
energy per nucleon of about 2.4 MeV.
Here one can expect a sizable contribution from excited QPs undergoing
a fission-like breakup, with a wide range of charge asymmetries.
Finally, in the region of
intermediate mass fragments (IMF, with 3$\leq$Z$\alt$16), the velocity
distribution spreads out, spanning a wider range (especially towards lower
velocities, where it is more influenced by the detection threshold), 
while the most probable velocity value shows a weak increasing trend.
This region is likely populated by the already mentioned 
neck emissions or by very asymmetric, possibly non-equilibrated, 
fission-like processes \cite{casi93,stefa95,mcin10,wilcz10}).

Further insight into the reaction processes can be gained by looking at the 
inclusive charge distributions of Fig.~\ref{chargedis}, which  
have been normalized in the range 18$\leq$Z$\leq$28, where a fragment is
either a QP remnant or the heavier fragment of a binary split of the QP
(quasi-elastic fragments with Z$\geq$29 are not used because their yield 
is too sensitive to the small difference in grazing angle 
between the two reactions or to an exact alignment of the beam).
This normalization roughly corresponds to considering the same number of 
inelastic binary or quasi-binary events.
The comparison of the two distributions clearly shows that the n-poor 
system $^{84}$Kr+$^{112}$Sn produces appreciably more IMFs.
This is probably due to the fact that in the n-poor system 
the break-up into lighter fragments with Z$<$15, either by fission or 
fragmentation, is favored with respect to the n-rich system.

The good isotopic resolution of the telescope allows to investigate the 
isotopic composition of the fragments.
For each element from Z=3 to Z=20, Fig. \ref{fig_perc} shows the relative 
probability of observing the various isotopes in the collision of $^{84}$Kr
with $^{112}$Sn (red full dots) and $^{124}$Sn (black open dots).
Similarly to the C and Mg isotopes of Fig. \ref{fig_pi}(b) and (c),
one finds --for all fragments and not only for the lighter ones-- 
that the n-rich side is more populated for $^{124}$Sn than 
for $^{112}$Sn and, vice-versa, the n-poor side is more populated for 
$^{112}$Sn than for $^{124}$Sn.

\begin{figure}[t!]
\begin{center}
\includegraphics[width=8.5cm] {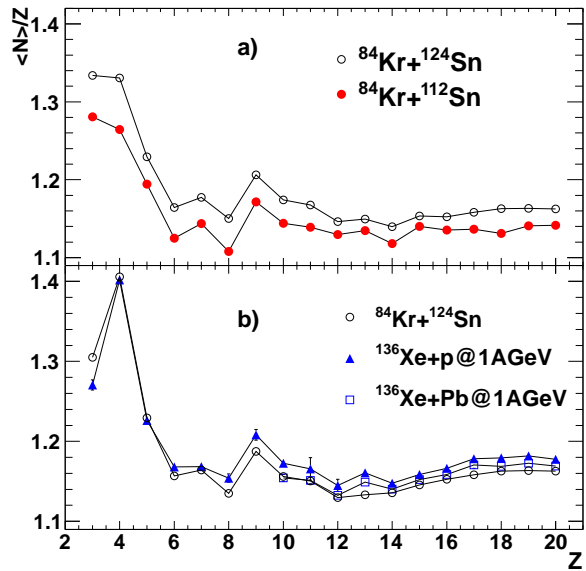}      
\end{center}
\caption{(Color online)
  (a) $\langle$N$\rangle$/Z as a function of Z
  for the reaction $^{84}$Kr+$^{124}$Sn
  (black open dots) and $^{84}$Kr+$^{112}$Sn (red full dots) at 35 MeV/u.
  Statistical errors are smaller than dot size.
  (b) Comparison of  $\langle$N$\rangle$/Z for 35 MeV/u $^{84}$Kr+$^{124}$Sn,
  (black open dots), 1 GeV/u $^{136}$Xe+H (blue triangles \cite{Nap2007})
  and 1 GeV/u $^{136}$Xe+Pb (open squares \cite{henzlova}),
  computed in the same common range of isotopes.}
\label{nsuz}
\end{figure}

A more quantitative estimate of the
different contributions of the two reactions to the n-rich 
and n-poor sides of the isotope distributions of Fig. \ref{fig_perc} 
is given by
the average number of neutrons per charge unit $\langle$N$\rangle$/Z.
This is an isospin sensitive variable which has been often used 
in the literature.
Values of $\langle$N$\rangle$/Z are shown in Fig. \ref{nsuz}(a) as a 
function of Z for the two collisions studied in this paper.
In the n-rich system
this ratio is systematically higher than in the n-poor one,
by an amount of about 0.03--0.05.
Since the largest part of the observed fragments belongs to the QP
region of the phase-space (see Fig. \ref{zetavel}), the observed difference
clearly demonstrates the action of an isospin diffusion mechanism:
the different isospin of the detected fragments depends on the 
n-richness of the target with which the projectile has interacted.

We can compare our results with published isotope-resolved 
cross sections (from mass spectrometer measurements at high 
bombarding energies), although some caution is required due to
our thresholds (see Fig. \ref{zetavel}).
The blue triangles of Fig. \ref{fig_perc} refer to published 
data \cite{Nap2007} for the spallation of 1 GeV/u $^{136}$Xe nuclei 
impinging on an Hydrogen target.
The data of \cite{Nap2007} are not very different from the results of 
this paper:
the main difference is that the abundance of isotopes on the
n-rich side of the distribution is further increased with respect to our
$^{124}$Sn (by an amount of the order of the difference between our two 
systems), and the abundance of isotopes on the n-poor side is further 
depressed with respect to our $^{112}$Sn (by about the same quantity).
The similarity with our results is quite surprising if one considers 
two facts:
first, spallation is a reaction mechanism completely different 
from that of our collisions and,
second, the isospin of $^{136}$Xe (N/Z=1.52) is considerably larger
not only than the $^{84}$Kr beam (1.33), but also than the
equilibrium value (1.42) of our system.
These observations suggest that the final fragment isospin content 
bears little dependence in the preceding dynamics,
but it retains memory of the original neutron richness.

In Fig. \ref{nsuz}(b), the isospin sensitive variable $\langle$N$\rangle$/Z
deduced from our n-rich system  
$^{84}$Kr+$^{124}$Sn (black open dots) is compared with that from 
the $^{136}$Xe spallation (blue triangles, \cite{Nap2007}) 
and the $^{136}$Xe+Pb collision (open squares, \cite{henzlova};
data available only for Z$\geq$10) at 1 GeV/u.
Because of incomplete isotopic distributions in some set of data 
(see, e.g., the lack of $^7$Be and of n-rich isotopes 
with Z=7--12 for the data of \cite{Nap2007} in Fig. \ref{fig_perc}),
a more meaningful comparison is obtained in Fig. \ref{nsuz}(b)
by computing $\langle$N$\rangle$/Z 
only from the isotopes that are common to the various sets of data.
Remarkably the lightest fragments of \cite{Nap2007} display a behavior very
similar to that of our data.
Above Z$\approx$8, the only significant difference is that the
fragments from the high-energy $^{136}$Xe reactions present just slightly
higher values of $\langle$N$\rangle$/Z 
with respect to the $^{84}$Kr+$^{124}$Sn reaction.
Similar differences with target isospin have been observed 
in the reactions $^{84}$Kr+$^{92,98}$Mo at 22 MeV/u \cite{lucas87}. 
However, those data are not included in Fig. \ref{nsuz}(b), 
because it is not specified which isotopes were detected. 
On the contrary, in the reactions $^{86}$Kr+$^{27}$Al, $^{103}$Rh, 
$^{197}$Au at 44 MeV/u, apparently no clear target dependence 
was observed \cite{bazin90}.

\begin{figure}[t!]
\begin{center}
\includegraphics[width=8.6cm] {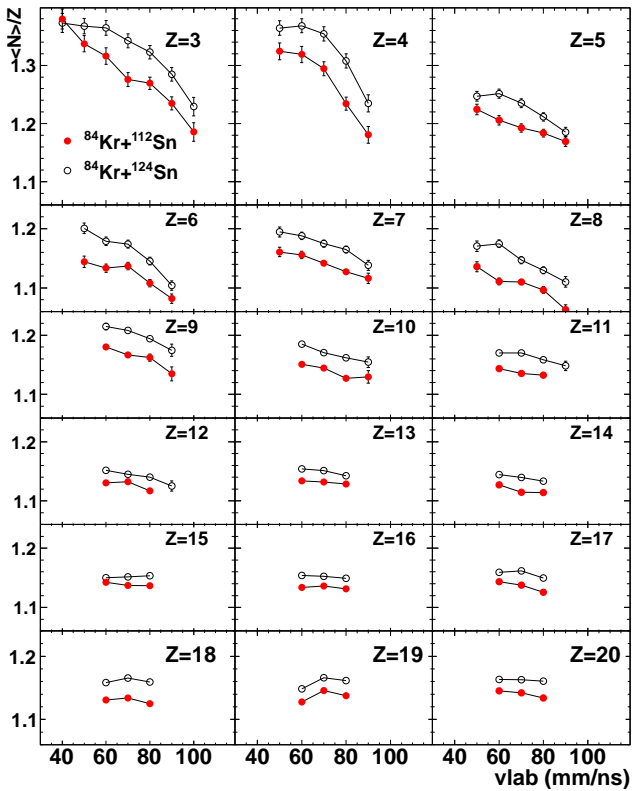}      
\end{center}
\caption{(Color online)
  $\langle$N$\rangle$/Z as a function of the laboratory velocity 
  for the reaction
  $^{84}$Kr+$^{124}$Sn (black points) and $^{84}$Kr+$^{112}$Sn (red points). 
  Each panel refers to a different element from $Z=3$ to $Z=20$.
  Error bars combine statistical errors and uncertainties in 
  the isotope identification.}
\label{nzsuzv}
\end{figure}

One may wonder whether there is a difference in the isospin content of the
fragments produced in the $^{84}$Kr+$^{124}$Sn and $^{84}$Kr+$^{112}$Sn
reactions, depending on the phase-space region they belong to.
For this purpose, Fig. \ref{nzsuzv} shows the evolution 
of $\langle$N$\rangle$/Z
for each element (from $Z=3$ up to $Z=20$) as a function of the 
laboratory velocity of the fragments.
The most evident effect is that, again, the black open dots 
(n-rich system)
are always above the red full dots 
(n-poor system).
This is an effect of the isospin diffusion, due to the interaction of
the projectile with targets of different isospin content.
The second clear observation is that for light ions $\langle$N$\rangle$/Z
rapidly decreases with increasing velocity, 
while it displays a rather flat behavior for heavier ions. 
The third point worth noting is that 
the highest values of $\langle$N$\rangle$/Z of
fragments with Z=3--4 are reached at the smallest laboratory velocities
(close to that of the center of mass).

Given the experiment geometry
(4.8$^\circ \leq \theta_\mathrm{lab} \leq 6^\circ$), 
the fragments with large velocities (of the order of that of the beam)
are likely to be emitted in forward direction 
from an excited QP, while those with lower velocities are expected to 
be emitted by the same QP in backward direction, 
with possible contributions from midvelocity- (or neck-) emissions.
In fact, at Fermi energies, fragments may be produced not only by a 
fission-like equilibrated decay of the QP (or QT), but also by the breakup 
of an elongated neck-like structure \cite{ditoro} formed between QP and QT.
It has been shown \cite{colin} that these fragments present
a kind of ``hierarchy effect'': 
lighter fragments originating from the thinner central part of the 
rupturing neck have small velocities in the center of mass frame,
while heavier fragments produced in thicker zones of the neck  
possess larger velocities, close to (but still smaller than) 
that of the QP (or QT).
Therefore, 
in this picture the low-velocity
lightest fragments (Z=3, 4 and partially 5) 
of Fig. \ref{nzsuzv} are probing the most central part of the neck 
and thus their higher
values of $\langle$N$\rangle$/Z could be an 
indication of isospin drift, namely a neutron enrichment of the 
more diluted central region of the neck \cite{baran05}.
On the contrary, heavier fragments with Z$\agt$12
have a low $\langle$N$\rangle$/Z (around 1.15) with practically 
no dependence on the emission velocities, which however span a rather narrow 
range of about 20-30 mm/ns around v$_\mathrm{lab} \approx$70 mm/ns.

\begin{figure}[t!]
\begin{center}
\includegraphics[width=8.4cm] {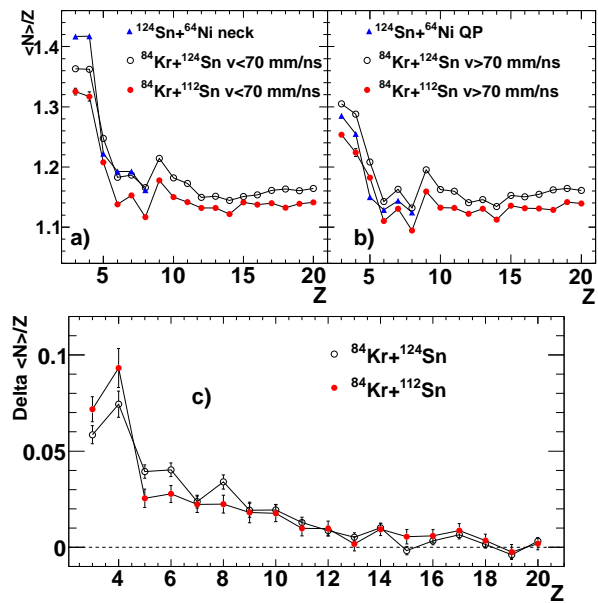}      
\end{center}
\caption{(Color online)
  $\langle$N$\rangle$/Z as a function of Z in the 
  n-rich (open black dots) and n-poor system (full red dots), for
  (a) v$_\mathrm{lab}<$ 70 mm/ns and
  (b) v$_\mathrm{lab}\geq$ 70 mm/ns. 
  Blue triangles: results of reference \cite{defili12}.
  (c) Differences between backward and forward values of 
  $\langle$N$\rangle$/Z for the two reactions of this paper.}
\label{nsuzv2}
\end{figure}

It is interesting to compare the data of Fig. \ref{nzsuzv} with the results
\cite{defili12} of the similar system $^{124}$Sn+$^{64}$Ni at 35 MeV/u.
In \cite{defili12} the assumed neck emissions and the 
more equilibrated decays of the QP have been selected on the basis 
of an angular correlation of the observed fragments. 
In our case, since we have only a single detected fragment, 
the selection is made on the basis of the laboratory velocity.
In figure \ref{nsuzv2}, the ratio $\langle$N$\rangle$/Z as a function of Z
is presented for v$_\mathrm{lab} <$ 70 mm/ns (a) 
and v$_\mathrm{lab} \geq$ 70 mm/ns (b) 
for the two systems measured in this work
(open black and full red dots for the n-rich and n-poor system, 
respectively).
These two selections on v$_\mathrm{lab}$ roughly correspond to 
light fragments emitted in backward and forward direction in the
frame of a QP.
The blue triangles (available only for 3$\leq$Z$\leq$8)
are the data of \cite{defili12}, ascribed (a) to 
neck emissions or (b) to QP emissions.
Although the selections are not exactly the same, 
the remarkable agreement with our data supports the interpretation 
that low-velocity light fragments are emitted from the neck.

One may further note that, in both systems, our data show no
appreciable forward-backward effect for fragments above Z$\approx$10.
This is better seen from Fig. \ref{nsuzv2}(c), which
shows the difference 
($\langle$N$\rangle$/Z)$_\mathrm{backw}$ -
($\langle$N$\rangle$/Z)$_\mathrm{forw}$
for the n-rich and n-poor systems.
Here the
effects of the isospin diffusion mechanism, 
which in each system  affects in the same way the forward- and 
backward-emitted isotopes, cancel out.
Thus the positive signal that is apparent for the light 
fragments has to be considered a signature of isospin drift.

\section{SUMMARY AND CONCLUSIONS}

We have presented data collected by the FAZIA collaboration
during a test experiment with a setup of small solid angle, 
but of high quality performances in terms of isotopic resolution 
(up to Z=20) for the systems $^{84}$Kr+$^{112}$Sn
and $^{84}$Kr+$^{124}$Sn at 35 MeV/u.  

The angular geometry of the setup (located close to the grazing angles 
for both reactions) allows to detect products originating from the 
quasi-projectile decay (including the quasi-projectile residue itself) 
and also from a phase-space region (close to the center of mass of the system)
where a sizable contribution of light ions produced in the neck-zone 
is expected.

Even with this simple setup, one can obtain significant information on 
isospin transport processes.
For each element, the relative yields show an enhancement of n-rich isotopes 
for the interaction of the $^{84}$Kr with a $^{124}$Sn target, 
and vice-versa an enhancement of n-poor isotopes for the interaction 
with a $^{112}$Sn target.
The fact that fragments emitted from the QP display a different 
neutron enrichment depending on the different isospin content of the
targets is a direct evidence of an isospin diffusion effect,
i.e. the transport of nucleons between projectile and target 
with different N/Z during the interaction phase.  
The relative yields are quite similar to those obtained in
\cite{Nap2007} for the spallation of $^{136}$Xe,
i.e. in a completely different scenario from the point of view of the 
reaction mechanism. 
The data show no appreciable dependence on the dynamics of the reaction.
A signature of the previous history remains only in the
differences in neutron richness associated with the different targets
and the same $^{84}$Kr beam.

According to theoretical studies, the neck region should be diluted with 
respect to the normal nuclear density.
On these grounds, an isospin drift is expected, which tends to increase 
the neutron richness of the neck region.
This prediction appears to be confirmed by the present data.
Light fragments, emitted in a possibly diluted phase-space region 
close to the center of mass of the system, display indeed 
a higher $\langle$N$\rangle$/Z, which strongly decreases when moving away 
from the neck region, towards the larger velocities typical of 
the decays of an excited QP.

On the contrary, heavier fragments (with Z$\agt$12) do not show any 
dependence of their $\langle$N$\rangle$/Z on the velocity bin; 
this fact can be 
understood by assuming that heavier fragments originate from the 
quasi-projectile fission, i.e. they have all a common origin, 
independently of their laboratory velocity (which spans a
considerably smaller range with respect to light fragments).
The $\langle$N$\rangle$/Z associated to the n-poor system is always 
smaller than that associated to the n-rich system for all velocity bins.

The investigation of isospin transport needs further experiments
and it will certainly benefit from the new facilities for radioactive beams 
now under construction 
and from large area multidetectors with A and Z 
identification, like FAZIA.
In fact these phenomena will be strongly enhanced if the difference of 
isospin content between the interacting nuclei can be further increased.

\begin{acknowledgments}
The authors would like to thank the crew of the Superconducting Cyclotron,
in particular D. Rifuggiato, for providing a very good
quality beam, and the staff of LNS for continuous support.
The support of the detector and mechanical workshops of the Physics Department 
of Florence is also gratefully acknowledged.
The research leading to these results has received funding from the European
Union Seventh Framework Programme FP7(2007-2013) under Grant Agreement No.
262010-ENSAR.
We acknowledge support by the Foundation for Polish Science - MPD program,
co-financed by the European Union within the European Regional Development 
Fund.
\end{acknowledgments}

\end{document}